\documentclass[11.5pt,a4paper]{article}
\usepackage[margin=2.5cm]{geometry}
\usepackage[utf8]{inputenc}
\usepackage[T1]{fontenc}
\usepackage{lmodern}

\usepackage{amsmath,amssymb,amsthm,bm}
\usepackage{graphicx}
\usepackage{float}
\usepackage{tabularx}
\usepackage{booktabs}
\usepackage{authblk}

\usepackage[ruled,vlined,linesnumbered]{algorithm2e} 
\SetKwFunction{FCheck}{CheckInput}
\SetKwFunction{FComplete}{CompleteInput}
\SetKwFunction{FSeq}{ExtractSequentialInput}
\SetKwFunction{FStruct}{ExtractStructuredInput}

\usepackage[svgnames,usenames]{xcolor}
\definecolor{myblue}{rgb}{0.21, 0.34, 0.74}
\definecolor{myred}{rgb}{0.79, 0.0, 0.09}
\definecolor{mygreen}{rgb}{0, 0.32, 0}
\colorlet{darkgreen}{ForestGreen}

\definecolor{matlabblue}{RGB}{0,0,180}
\definecolor{matlabviolet}{RGB}{167,9,245}
\definecolor{matlabgreen}{RGB}{0,128,19}
\definecolor{pythonteal}{RGB}{0,120,140}

\usepackage{listings}
\lstdefinestyle{MatlabStyle}{
    language=Matlab,
    basicstyle=\ttfamily\small,
    keywordstyle=\color{matlabblue},
    commentstyle=\color{matlabgreen},
    stringstyle=\color{matlabviolet},
    alsoletter={._},                     
    morekeywords={whos,validateattributes}, 
    numbers=none,
    showstringspaces=false,
    breaklines=true,
    frame=none,                   
    backgroundcolor=\color{gray!5},  
    escapeinside={(*@}{@*)}
}
\lstdefinestyle{PythonStyle}{
    language=Python,
    basicstyle=\ttfamily\small,
    keywordstyle=\color{pythonteal},
    commentstyle=\color{yellow!50!black},
    stringstyle=\color{red},
    alsoletter={._},            
    morekeywords={time,time.time,np.dot,np,range,len,np.any,flatten,np.isfinite,np.finfo,np.float64,.eps}, 
    numbers=none,
    showstringspaces=false,
    breaklines=true,
    frame=none,                   
     backgroundcolor=\color{gray!5},  
    escapeinside={(*@}{@*)}
}

\usepackage{soul}
\usepackage{cancel}
\usepackage{fontawesome}

\usepackage{xcolor}
\definecolor{DeepTeal}{RGB}{0, 102, 102}   
\definecolor{SlateBlue}{RGB}{0, 92, 175}   
\definecolor{ElectricCyan}{RGB}{0, 191, 255} 
\usepackage[colorlinks=true]{hyperref}
\hypersetup{
    linkcolor=ElectricCyan,      
    citecolor=ElectricCyan,     
    urlcolor=cyan,           
    filecolor=teal,          
    menucolor=DeepTeal,
    runcolor=ElectricCyan
}

\usepackage{orcidlink}

\newtheorem{Theorem}{Theorem}

\title{EPITIME: A Computational Framework for Integral Epidemic Models with Structure-Preserving Discretizations}

\author[1,\orcidlink{0000-0003-4998-3725}]{Bruno Buonomo\thanks{Corresponding Author: bruno.buonomo@unina.it. The authors contributed equally to this work.}}
\author[1,\orcidlink{0000-0003-4545-6266}]{Eleonora Messina}
\author[1,\orcidlink{0009-0007-4390-1449}]{Claudia Panico}
\author[2,\orcidlink{0000-0002-1869-945X}]{Mario Pezzella}
\author[3,\orcidlink{0000-0002-5282-9158}]{Gaetano Zanghirati}

\affil[1]{\small University of Naples Federico II, Department of Mathematics and Applications ``Renato Caccioppoli'', Via Cintia, Naples, 80126, Italy}
\affil[2]{\small National Research Council of Italy, Institute for Applied Mathematics ``Mauro Picone'', Via P. Castellino, Naples, 80131, Italy}
\affil[3]{\small University of Ferrara, Department of Mathematics and Computer Science, Via Saragat, Ferrara, 44122, Italy}

\date{} 

\begin{document}

\maketitle

\begin{abstract}
We present EPITIME (EPidemic Integral models TIMe profile Explorer), a computational framework for the simulation of two classes of integral epidemic models: an age of infection model and an information dependent behavioural model.
The framework combines structure preserving Non-Standard Finite Difference discretizations with modular implementations in MATLAB and Python, together with routines for parameter handling, input validation, performance assessment, and graphical interaction.
The proposed methods preserve key qualitative properties of the continuous problems, including positivity, boundedness, invariant regions, and correct long term behaviour, independently of the time step.
We outline the numerical schemes for both model classes and their main analytical properties, including first order convergence.
We then describe the software architecture and illustrate its use through numerical experiments on asymptotic behaviour, inverse reconstruction of an infectivity kernel from COVID 19 incidence data, and behavioural dynamics under different memory kernels.
Overall, EPITIME provides a reliable and accessible computational environment for the numerical study of renewal epidemic models.
\end{abstract}

\vspace{0.5cm}


\section{Introduction}
Mathematical epidemiology has long relied on compartmental models to describe the spread of infectious diseases, with SIR- and SEIR-type systems providing a standard framework for the analysis of transmission dynamics, threshold quantities and control strategies \cite{BrauerBook, Martcheva_2015}.  However, several epidemiologically relevant mechanisms, including infection-age dependence, variable infectiousness and memory effects, are more naturally represented within integral formulations than through finite-dimensional compartmental systems alone \cite{Kermack1927}. Integral models replace constant transition rates with infectivity kernels, allowing for a more general representation of how infectiousness evolves over the course of the disease and naturally incorporating distributed delay effects, such as those induced by incubation and latency periods.

In this context, Brauer and coauthors \cite{BrauerBook} provided comprehensive treatments of  \emph{Age-of-Infection} (AoI) epidemic models, clarifying their analytical properties and their connections to standard compartmental systems. Furthermore, Diekmann and collaborators developed a unified renewal-equation framework for epidemic models that systematically incorporates key biological features such as demography, non-permanent immunity and heterogeneous infectivity. Within this setting, the Force of Infection (FoI), that is the rate at which susceptibles become infected, is expressed through a scalar renewal equation driven by an infectivity function. Classical compartmental models are recovered as particular limiting cases \cite{JBD2012, Diekmann2018, Diekmann2025}.  

In parallel, increasing attention has been devoted to epidemic models that account for behavioural adaptation, where transmission is affected not only by biological factors but also by changes in human behaviour driven by information, perceived risk, or epidemic awareness \cite{ManfrediDonofrio2013}.  In this setting, integral formulations are particularly natural, since both infection-age effects and behavioural responses can be described through history-dependent terms and distributed kernels. This perspective is especially relevant for information-dependent epidemic models, in which transmission is modulated by an information index reflecting the effect of past epidemic trends on present contact behaviour \cite{bai2014,BehRE_jmb, Bumi2025,donofrio2009}.

The numerical approximation of integral epidemic models poses several challenges, especially when one aims to preserve crucial qualitative features of the continuous dynamics, such as positivity, boundedness, invariant regions, equilibrium structure and long-time behaviour. In epidemiological applications, the loss of these structural properties may lead to spurious numerical behaviours and unreliable long-time simulations. For this reason, Non-Standard Finite Difference (NSFD) methods and other structure-preserving discretization techniques have become increasingly relevant in the numerical analysis of epidemic models \cite{Mickens2020_NSFD, NSFD_MPV,Lubuma_2014}. These methods are specifically designed to retain key dynamical properties of the underlying continuous problem independently of the time step, thus providing a more robust basis for scientific computation.

From a computational point of view, while compartmental epidemic models formulated as ODE systems can rely on widely available and well-established software environments, software for epidemic models governed by integral equations is much less common. This gap is particularly relevant because the specific structure of integral epidemic models, with history-dependent terms, often requires ad hoc implementations rather than the direct use of standard-purpose solvers. In this setting, dedicated computational frameworks that combine usability, reproducibility and numerical robustness are especially valuable. This approach is in line with ongoing work on reliable software for dynamical systems, such as the tools provided by the CDLab (Computational Dynamics Laboratory), University of Udine, for the numerical analysis of delay equations and related models \cite{CDLabSoftware}.

Motivated by these considerations, in this work we present EPITIME (EPidemic Integral models TIMe-profile Explorer), a computational framework for the simulation of two classes of integral epidemic models: an age-of-infection model and an information-dependent behavioural model. The proposed package combines structure-preserving NSFD discretizations with modular software implementations in both MATLAB and Python. To this end, EPITIME includes validated solver modules, routines for parameter handling and consistency checks, performance indicators and a graphical user interface designed to facilitate simulations for users with limited familiarity with MATLAB or Python programming. 
This design makes the framework suitable for a variety of tasks, including qualitative dynamical investigations, numerical validation of theoretical properties and reliable long-time simulations. Overall, the proposed framework is intended to bridge the gap between the theoretical study of integral epidemic models and their practical implementation.

The paper is organized as follows. In Section \ref{sec:Problem_Formulation} we present the epidemic models under consideration and discuss a general framework including both formulations. In Section \ref{sec:Solutions and Algorithms} we review the corresponding NSFD discretizations and their main qualitative properties. Section \ref{sec:Tool_Description} is devoted to the description of the EPITIME software environment, including its MATLAB and Python implementations and the graphical user interface. In Section \ref{sec:Numerical_Experiments} we report a selection of numerical experiments, including tests on asymptotic behaviour, inverse-problem reconstruction and simulations illustrating the qualitative dynamics of the behavioural model. Final remarks are given in Section \ref{sec:Conclusions}.

\section{Model Formulation}\label{sec:Problem_Formulation}
In the following sections, we focus on different classes of integral epidemic models, distinguishing between formulations in which the FoI depends solely on biological factors and those that explicitly incorporate human behavioural feedback through information-dependent transmission mechanisms.

\subsection{The Age-of-Infection epidemic model}\label{sec:Original_AoI}
The model defined in \cite[Chapter 4.5]{BrauerBook} traces the evolution of an epidemic within a closed population taking into account the cumulative contribution of past infections to the current FoI. The model reads
\begin{equation}\label{eq:AoI_Infectivity}
    \begin{split}
        S^{\prime}(t)&=-\beta_0 S(t)\varphi(t),\\
        \varphi(t)&=\beta_0\int_{0}^{+\infty} A(\tau) S(t-\tau)\varphi(t-\tau)\, d\tau, 
    \end{split}
\end{equation}
and we refer to Table \ref{tab:Mario} for a complete description of the variables, parameters and functions.
\begin{table}[H]
  \caption{Functions and parameters of the model \eqref{eq:AoI_Infectivity}.}
  \label{tab:Mario}
  \newcolumntype{Y}{>{\centering\arraybackslash}X}
  \begin{tabularx}{\textwidth}{c Y Y}
  \toprule
  \textbf{Unknowns} & \textbf{Meaning} & \textbf{Properties} \\
  \midrule
  $S(t)$ & Size of susceptible compartment at time $t$ &
  $0<S(\infty)\leq S(t) \leq S_0$ and $S^\prime(t)<0,$ $\forall t\geq 0.$\\[0.4cm]
  $\varphi(t)$ & Total infectivity of members of the population at time $t$ &
  $\varphi(t)\geq 0,$ $\forall t\geq 0$ and $\lim\limits_{t\to\infty}\varphi(t)=0.$\\[0.35cm]
  \toprule
  \textbf{Functions} & \textbf{Meaning} & \textbf{Properties} \\
  \midrule
  $A(\tau)$ &
  Expected contribution to transmission dynamics of an individual whose infection age is $\tau$ &
  $A(\tau)\geq 0,$ $\forall \tau \geq 0$ and $A\in L^1(\mathbb{R}^+_0).$ \\[0.35cm]
  \toprule
  \textbf{Parameters} & \textbf{Meaning} & \textbf{Properties} \\
  \midrule
  $N$ & Total population size & $N>0.$\\[0.25cm]
  $S_0$ & Initial susceptible population & $S_0\in [0,N].$\\[0.25cm]
  $\beta_0$ & Rate of effective contacts & $\beta_0>0.$\\[0.25cm]
  $R_0$ & Basic reproduction number &
  $R_0 = \beta_0 N\int_0^{\infty}A(\tau)\, d\tau.$\\[0.25cm]
  $S(\infty)$ & Final size of the epidemic &
  $S(\infty)=\lim_{t\to\infty}S(t)\in (0,S_0].$\\[0.1cm]
  \bottomrule
  \end{tabularx}
\end{table}
Here, $\tau$ denotes the infection age, that is the time elapsed since the infection. For this reason, from now on, we refer to \eqref{eq:AoI_Infectivity} as the Age-of-Infection (AoI) model. An alternative expression in terms of the force-of-infection function is presented in \cite{Diekmann2025}, where also the connection to compartmental models is explored (see \cite[Sec. 9.3]{Diekmann2018} and \cite[Sec. 4 and Sec. 6]{Diekmann2023} for further details). Moreover, a broader framework, which incorporates symptomatic and asymptomatic infections \cite{Bai2023} as well as heterogeneously mixed populations \cite{BrauerWatmough2009,Mixing_MPV}, is discussed in \cite{Elefante}.

A numerical approximation method for \eqref{eq:AoI_Infectivity}, specifically designed to unconditionally preserve the qualitative properties of its solution, is recalled in Section \ref{subsec:NSFD_AoI}. Furthermore, some insights on the asymptotic behaviour of the continuous and numerical solutions to \eqref{eq:AoI_Infectivity} are provided in Section \ref{subsec:AoI_Asympt}.

\subsection{The integral behavioural epidemic model}
Here, we consider an integral epidemic model that accounts for changes in individual behaviour during an epidemic outbreak by incorporating a time varying information index $M(t)$. Specifically, this function depends on current and past epidemiological quantities and describes how individuals adjust their risky contacts in response to information and rumours about the disease. The model, whose formulation and analysis are detailed in \cite{BehRE_jmb}, reads 
\begin{equation}\label{model_beh}
    \begin{aligned}
        S'(t) &= \lambda - \mu S(t) - \beta(M(t)) S(t) F(t),\\
        F(t) &= \int_0^{\infty} A_{\mu}(\tau)\, \beta(M(t-\tau))\, S(t-\tau)\, F(t-\tau)\, d\tau,\\
        M(t) &= \int_0^{\infty} K(\tau)\, H(t-\tau)\, g(F(t-\tau))\, d\tau.
    \end{aligned}
\end{equation}

\noindent Table \ref{tab:Claudia} provides the unknowns, parameters and known functions of model \eqref{model_beh}, together with their interpretations and the associated assumptions. Some of the key functions involved are described in more detail in what follows. 

In particular, the function \(A_{\mu}(\tau)=e^{-\mu \tau}A(\tau)\) describes the infectivity profile over time, where \(e^{-\mu \tau}\) accounts for demographic mortality.  
The inhibition term \(\beta(M(t))\)  captures the reduction in transmission due to information and rumours about the disease. The quantity $\beta(M(t)) F(t)$ represents the FoI at time $t$. The message function $g$ and the memory kernel $K$ represent the perception of infection risk and how past information influences current behaviour, respectively. Finally, $H(\cdot)$ is the Heaviside step function.

Model \eqref{model_beh} can be regarded as a natural extension of the work by Breda et al. \cite{JBD2012}, starting from their formulation and incorporating a dependence of the FoI on the information index.
A variant of \eqref{model_beh} including incidence--dependent contact patterns is outlined in \cite{Bumi2025}. There, $M(t)$ is expressed as a weighted sum of past and present contributions of incidence trends, given by $k\beta(M(t))S(t)F(t)$, where $k \in (0,1)$ represents the information coverage.

\begin{table}[H]
  \caption{Functions and parameters of the model \eqref{model_beh}.}
  \label{tab:Claudia}
  \newcolumntype{Y}{>{\centering\arraybackslash}X}
  \begin{tabularx}{\textwidth}{c Y Y}
  \toprule
  \textbf{Unknowns} & \textbf{Meaning} & \textbf{Properties} \\
  \midrule
  $S(t)$ & Size of the susceptible compartment at time $t$ &
  $0< S(t) \leq S_{max} < +\infty$,   $\forall t\geq 0.$ \\[0.4cm]
  $F(t)$ & Driving factor of the FoI at time $t$ &
  $ 0 \leq F(t) \leq F_{max}< + \infty,$ $\forall t\geq 0.$ \\[0.4cm]
  $M(t)$ & Information index at time $t$ &
  $ 0 \leq M(t) \leq M_{max}<+\infty,$ $\forall t\geq 0.$ \\[0.35cm]
  \toprule
  \textbf{Functions} & \textbf{Meaning} & \textbf{Properties} \\
  \midrule
  $A$ & Infectivity function & $A(\tau) \geq 0, \forall \tau \geq 0$ and $A(\tau) \in L^1(\mathbb{R}^+_0).$ \\[0.25cm]
  $A_{\mu}$ & Infectivity function with demography & $A_{\mu}(\tau) \geq 0,$ $\forall \tau \geq 0$ and $A_{\mu}(\tau) \in L^1(\mathbb{R}^+_0)$. \\[0.5cm]
  $\beta$ & Inhibition function & $\beta(0)=1$, $\beta(M)>0$, $\beta'(M) < 0.$ \\[0.25cm]
  $g$ & Message function & $g(0)=0$, $g(F) \geq 0$, $g^{\prime}(F) > 0.$ \\[0.25cm]
  $K$ & Memory kernel & $K(\tau) >0,$ $\forall \tau \geq 0,$ \ $K(\tau) \in L^1(\mathbb{R}^+_0)$ and $\|K\|_{L^1(\mathbb{R}^+_0)}=1$. \\[0.25cm]
  \toprule
  \textbf{Parameters} & \textbf{Meaning} & \textbf{Properties} \\
  \midrule
  $\lambda$ & Net inflow of susceptibles & $\lambda>0.$\\[0.25cm]
  $\mu$ & Natural death rate & $\mu>0.$\\[0.25cm]
  $N$ & Total population size & $N=\frac{\lambda}{\mu}>0.$\\[0.25cm]
  $S_0$ & Initial susceptible population & $S_0\in [0,N].$\\[0.25cm]
  $R_0$ & Basic reproduction number &
 $R_0 = N\int_0^\infty A_\mu(\tau)\,d\tau$\\[0.25cm]
  \bottomrule
  \end{tabularx}
\end{table}

Section \ref{subsec:NSFD_BEH} reviews a numerical approximation technique for the model \eqref{model_beh} that is specifically constructed to unconditionally preserve the qualitative properties of its solution.

\section{Solution Methods and Algorithms}\label{sec:Solutions and Algorithms}

In order to discuss efficient simulation methods for the models introduced in the previous sections, we consider a unifying $(d+1)$--dimensional framework that generalizes both \eqref{eq:AoI_Infectivity} and \eqref{model_beh}. To this end, we denote by $S(t)\in \mathbb{R}^+$ and $X(t)=[X_1(t),\dots,X_d(t)]^\mathsf{T}\in\mathbb{R}^{d}$ the unknown functions, and by $Q(t,\tau)\in\mathbb{R}^{d \times d}$ the prescribed kernel of the problem. More specifically, we address the system
\begin{equation}\label{eq:unico_modello}
    \begin{split}
        & S'(t)=\alpha_1 - S(t)\left(\alpha_2 + G_1(X(t))\right), \\
        & X(t)=\int_0^{+\infty} Q(t,\tau)G_2(S(t-\tau),X(t-\tau)) \, d\tau,
    \end{split}
\end{equation}
where the constants $\alpha_1,\alpha_2\in \mathbb{R}^+_0$, and the non-linear functions $G_1:\mathbb{R}^{d}\to \mathbb{R}$ and $G_2:\mathbb{R}^{d+1}\to \mathbb{R}^{d}$ are given. In particular, the formulations \eqref{eq:AoI_Infectivity} and \eqref{model_beh} are recovered from \eqref{eq:unico_modello} as special cases corresponding to the choices. For $d=1$:
\begin{equation*}
    \begin{split}
        & X(t)=\varphi(t), \quad Q(t,\tau)=A(\tau), \qquad
        G_1(X)=\beta_0 X, \quad G_2(S,X)=G_1(X)S,
        \end{split}
        \end{equation*}
        and, for $d=2$:
        \begin{equation*}
        \begin{split}
        & X(t)=\begin{bmatrix} F(t) \\ M(t) \end{bmatrix}, \;
        Q(t,\tau)=\begin{bmatrix} A_\mu(\tau) & 0 \\ 0 & \hspace{-0.3cm}H(t-\tau)K(\tau) \end{bmatrix}, \; G_1(X)=\beta(X_2)X_1, \; 
        G_2(S,X)=\begin{bmatrix} G_1(X)S \\ H(t)g(X_1) \end{bmatrix},
    \end{split}
\end{equation*}
respectively.

The numerical discretizations presented in the subsequent  sections are based on the following first-order non-standard approximations of the derivative and integral operators
\begin{equation}\label{eq:appox_continua}
    \begin{split}
        & S'(t+h)\approx \dfrac{S(t+h)-S(t)}{h \gamma_1(h)}
        \approx \alpha_1 - S(t+h)\left(\alpha_2 + G_1(X(t))\right),  \\
        & \int_{t}^{t+h} Q(\tilde{t},\tilde{t}-\tau)G_2(S(\tau),X(\tau)) \, d\tau
        \approx h\Gamma(h) \, Q(\tilde{t},\tilde{t}-t)G_2(S(t+h),X(t)),
    \end{split}
\end{equation}
which hold for each fixed $\tilde{t}\geq t+h$ and $h>0$, with $\Gamma(h)=\text{diag}\left(\gamma_2(h), \dots, \gamma_{d+1}(h)\right)$ and $\gamma_i(h)=1+\mathcal{O}(h),$ $i=1, \dots, d+1,$ positive functions. Specifically, \eqref{eq:appox_continua}$_1$ relies on a weighted forward finite difference, while \eqref{eq:appox_continua}$_2$ corresponds to a modified rectangular quadrature rule where $Q$ and $X(t)$ are evaluated at the left end-point and $S(t)$ at the right one. This choice is aimed at preserving, independently of the time step-size $h$, the qualitative properties of the underlying continuous system when passing to a discrete-time formulation. The corresponding linearly implicit numerical schemes for \eqref{eq:AoI_Infectivity} and \eqref{model_beh} are derived and discussed in detail in the following sections.

Throughout the paper, we assume that the properties of the functions in Tables \ref{tab:Mario} and \ref{tab:Claudia} hold true. Any additional condition needed for the analysis of the numerical asymptotic behaviour will be explicitly stated as required.

\subsection{The NSFD discretization for the Age-of-infection model}\label{subsec:NSFD_AoI}
Throughout this section, we consider problem \eqref{eq:AoI_Infectivity} and assume that the non-negative function
\begin{equation}\label{eq:phi0_Cond}
    \varphi_0(t)=\beta_0\int_{t}^{+\infty} A(\tau)\, S(t-\tau)\,\varphi(t-\tau)\, d\tau,
\end{equation}
is prescribed. As discussed in \cite{Brauer2008}, $\varphi_0(t)$ represents the total infectivity at time $t$ of individuals infected prior to the initial outbreak. In general, it satisfies
\begin{equation}\label{eq:Bound_phi0}
    0 \leq \varphi_0(t) \leq (N - S_0)\, A(t),
\end{equation}
with equality attained when all such individuals have infection age zero at $t=0$.

Let $ t_n = n h $, for $ n = 0, 1, \dots $, denote a uniform temporal grid with step-size $ h > 0 $, and let $[S_n, \ \varphi_n]^\mathsf{T}$ denote the discrete approximations of the solution to \eqref{eq:AoI_Infectivity} at time $ t_n $. The Non-standard Finite Difference (NSFD) scheme introduced in \cite{NSFD_MPV} is then given by 
\begin{equation}\label{eq:NSFDscheme}
    \begin{split}
        S_{n+1} &= S_n - h \beta_0 S_{n+1}\varphi_n, \\
        \varphi_{n+1} &= \varphi_0(t_{n+1}) + h \beta_0 \sum_{j=0}^{n} A(t_{n+1-j}) S_{j+1}\varphi_j,
    \end{split}
\end{equation}
for $ n = 0, 1, \ldots, $ with given initial values $ S_0 = S(0) $ and $ \varphi_0 = \varphi_0(0) $. The method \eqref{eq:NSFDscheme} relies on the non-local approximations outlined in \eqref{eq:appox_continua}, with $\gamma_i(h)=1.$ Its formulation gives rise to the straightforward Algorithm \ref{alg:AoI} that preserves the structure of the model with minimal computational effort.

\begin{algorithm}
\caption{Pseudo-code of the NSFD scheme \eqref{eq:NSFDscheme} for the Age-of-Infection model \eqref{eq:AoI_Infectivity}}\label{alg:AoI}
\DontPrintSemicolon
\KwIn{$S_0, \, N, \, T, \, h, \, \beta_0, \, A(\cdot), \, \varphi_0(\cdot)$}
\KwOut{$\bm{t}=[t_0,\dots,t_{N_t}]^{\mathsf{T}}, \ \bm{S}=[S_0,\dots,S_{N_t}]^{\mathsf{T}}, \ \bm{\varphi}=[\varphi_0,\dots,\varphi_{N_t}]^{\mathsf{T}} $}
$B\leftarrow \beta_0 h, \qquad \hat{B} \leftarrow  0, \qquad \nu_t \leftarrow \lfloor T/h\rfloor$\;
\tcp*{Grid construction and step adjustment}
\For{$n=0$ \KwTo $\nu_t$}{$t_n \leftarrow n h$}
\If{$\nu_t h < T$}{
    $\hat{B} \leftarrow \beta_0(T - \nu_t h)$\;
    $t_{\nu_t + 1} \leftarrow T$\;
}
$N_t\leftarrow\texttt{length}(\bm{t}), \qquad \bm{\Phi}^0\leftarrow\varphi_0(\bm{t}), \qquad$ $\bm{\mathcal{A}}\leftarrow A(\bm{t})$ \;
\tcp*{Main time-stepping loop}
\For{$n = 0$ \KwTo $N_t-1$}{
    \If{$(\hat{B}\neq 0 \ \land \ n= N_t-1)$}{$B\leftarrow\hat{B}$}
    $S_{n+1} \leftarrow S_n/(1 + B \varphi_n)$\;
    $\varphi_{n+1} \leftarrow \Phi^0_{n+1} + B\displaystyle \sum_{j=0}^{n} \mathcal{A}_{n+1-j} \, S_{j+1} \, \varphi_j$
    } 
\Return $\bm{t}, \bm{S}, \bm{\varphi}$
\end{algorithm}

The method \eqref{eq:NSFDscheme} belongs to the class of NSFD methods, originally developed for differential equations (see \cite{Mickens2020_NSFD} and references therein) and recently extended to integral formulations \cite{Elefante,Lubuma_2014,Pezzella_ESAIM}. The following result establishes its linear convergence in case of finite-time integration. For a detailed error analysis we refer the reader to \cite[Lemma 3.1, Theorem 3.2 and Theorem 4.1]{NSFD_MPV}.
\begin{Theorem}\label{tmh:Conv-NSFD}
    Consider problem \eqref{eq:AoI_Infectivity} and assume that $A(t)\in C^1([0,T]),$ with $0<T<+\infty$ and $T=\bar{n}h.$ Denote by $E(h;t_n)=[S(t_n), \ \varphi(t_n)]^\mathsf{T}-[S_n, \ \varphi_n]^\mathsf{T}$ the global discretization error associated to the NSFD scheme \eqref{eq:NSFDscheme}. Then, 
    \begin{equation*}
        \max_{0\leq n \leq \bar{n}}\|E(h;t_n)\|=\mathcal{O}(h), \ \ \text{as} \ h\to0.
    \end{equation*}
\end{Theorem}

A key advantage of the NSFD discretization \eqref{eq:NSFDscheme} is that it unconditionally preserves the positivity, the monotonicity and the asymptotic behaviour of the continuous solution to \eqref{eq:AoI_Infectivity}. These properties, established in \cite[Theorem 3.4]{NSFD_MPV}, are summarized in the following result.
\begin{Theorem}\label{THMmimimcNSFD}
    Let $[S_n, \ \varphi_n]^\mathsf{T}$ be the solution to the discrete equation \eqref{eq:NSFDscheme} with initial values $S_0>0$ and $\varphi_0\geq0$. Then, independently of the step-size $h>0,$ the sequence $\{S_n\}_{n \in \mathbb{N}_0}$ is positive, non-increasing and bounded from above by $S_0.$ Therefore, there exists $\lim_{n\to \infty}S_n=S_{\infty}(h)\in(0,S_0].$ Moreover, $\{\varphi_n\}_{n \in \mathbb{N}_0}$ is a non-negative, bounded and vanishing sequence.
\end{Theorem}

\subsection{The non-local scheme for the integral behavioural epidemic model}\label{subsec:NSFD_BEH}
To solve the integro-differential system \eqref{model_beh}, we assume that the forcing terms
\begin{equation}\label{funz_iniz_short}
    F_0(t) = \int_{-\infty}^0 A_{\mu}(t-\tau) \beta(M(\tau)) S(\tau) F(\tau) d\tau, \quad
    M_0(t) = \int_{-\infty}^0 K(t-\tau) H(\tau) g(F(\tau)) d\tau,
\end{equation}
are given non-negative functions. More precisely, we set
\begin{equation*}
    F_0(t)=A_\mu(t)(\lambda-\mu S_0), \quad
    M_0(t)=g(F(0))K(t).
\end{equation*}
Discretizing the system \eqref{model_beh} as detailed in \eqref{eq:appox_continua}, with $\gamma_1(h)=\gamma_2(h)=1$ and $\gamma_3(h)=\gamma(h)$ to be specified later, yields the numerical method
\begin{equation}\label{mod_discreto_short}
\begin{aligned}
S_{n+1} &= S_n + h \big( \lambda - \mu S_{n+1} - \beta(M_n) S_{n+1} F_n \big),\\
F_{n+1} &= F_0(t_{n+1}) + h  \sum_{j=0}^{n} A_{\mu}(t_{n+1-j}) \beta(M_j) S_{j+1} F_j,\\
M_{n+1} &= M_0(t_{n+1}) + h  \gamma(h) \sum_{j=0}^{n} K(t_{n-j}) g(F_j),
\end{aligned}
\end{equation}
where \(S_n, F_n \) and \(M_n\) approximate \(S(t_n)\), \(F(t_n)\) and \(M(t_n)\) at the grid point \(t_n = n h\), with $h>0$ uniform step-size. Here, the initial values \(S_0 = S(0)\), \(F_0 = F(0) = F_0(0)\), and \(M_0 = M(0) = M_0(0)\) are given. A pseudo-code of the scheme \eqref{mod_discreto_short} is presented in Algorithm \ref{alg:BEH}.

\begin{algorithm}
\caption{Pseudo-code of the NSFD scheme \eqref{mod_discreto_short} for the behavioural epidemic model \eqref{model_beh}}
\label{alg:BEH}
\DontPrintSemicolon
\KwIn{$T,\, h,\, N,\, \mu, \, S_0,\,
A_{\mu}(\cdot),\, K(\cdot),\, \beta(\cdot), \, g(\cdot) $}
\KwOut{$\bm{t}=[t_0,\dots,t_{N_{t}}]^{\mathsf{T}},\;
\bm{S}=[S_0,\dots,S_{N_{t}}]^{\mathsf{T}},\;
\bm{F}=[F_0,\dots,F_{N_{t}}]^{\mathsf{T}},\;
\bm{M}=[M_0,\dots,M_{N_{t}}]^{\mathsf{T}}$}
$\nu_{t} \leftarrow \lfloor T/h \rfloor,\qquad D \leftarrow h \qquad \hat{D} \leftarrow 0$\;
\tcp*{Grid construction and step adjustment}
\For{$n=0$ \KwTo $\nu_t$}{$t_n \leftarrow n h$}
\If{$\nu_t h < T$}{
    $\hat{D} \leftarrow T - \nu_t h$\;
    $t_{\nu_t + 1} \leftarrow T$\;
}
\tcp*{initial values}
$N_{t} \leftarrow \texttt{length}(\bm{t}), \quad\bm{A^{\mu}} \leftarrow A_{\mu}(\bm{t}),\quad
\bm{K} \leftarrow K(\bm{t}), \quad \bm{F}^0 \leftarrow \mu (N-S_0)\bm{A^{\mu}},\quad \bm{M}^0 \leftarrow g(F_0)\,\bm{K}$\;
\tcp*{NSFD normalization factor}
Compute $\gamma$ as defined in \eqref{eq: factor gamma}\;
\tcp*{Main time-stepping loop}
\For{$n=0$ \KwTo $N_{t}-1$}{
  \If{$(\hat{D}\neq 0 \ \land \ n= N_t-1)$}{$D\leftarrow\hat{D}$}
    $S_{n+1} \leftarrow
    \dfrac{S_n+ \lambda D}
    {1+D(\mu+\,\beta(M_n)\,F_n)}$\;
    $F_{n+1} \leftarrow F^0_{n+1} + D \displaystyle\sum_{j=0}^{n} A^\mu_{\,n+1-j}\, S_{j+1}\,\beta(M_j)\,F_j$\;
    $M_{n+1} \leftarrow M^0_{n+1} + \gamma D \displaystyle\sum_{j=0}^{n} K_{\,n-j}\, g(F_j)$
\;
}
\Return $\bm{t},\,\bm{S},\, \bm{F},\,\bm{M}$
\end{algorithm}

A comprehensive analysis of the discrete system \eqref{mod_discreto_short} is carried out in \cite{Beh_discr_etna}, where the presence and the stability of discrete equilibria are investigated, as well. In what follows, we present only a brief overview of these findings, starting from the linear convergence of the method \eqref{mod_discreto_short}.


\begin{Theorem}\label{Th converg met}
    Consider the problem \eqref{model_beh} and assume that $A_{\mu}, K \in C^1([0,T])$, with $0<T<+\infty$ and $T=\bar{n}h.$ Denote by $E(h;t_n)=[S(t_n), \ F(t_n), \ M(t_n)]^\mathsf{T}-[S_n, \ F_n, \ M_n]^\mathsf{T}$ the global discretization error of the scheme \eqref{mod_discreto_short}. Then, 
    \begin{equation*}
         \max_{0\leq n \leq \bar{n}}\|E(h;t_n)\|=\mathcal{O}(h), \  \ \text{as} \ h\to0.
    \end{equation*}
\end{Theorem}

Assume, in addition to the properties listed in Table \ref{tab:Claudia}, that $A_{\mu}'(\tau), \, K'(\tau) \in L^1(\mathbb{R}_0^+)$ and define, for $h\in \mathbb{R}^+$, the functions
\begin{equation}\label{eq: discr Lapl transf}
\Bar{A}_{\mu}(h):=h\sum_{n=1}^{\infty}A_{\mu}(t_n),
\qquad
\Bar{K}(h):=h\, \gamma(h) \,\sum_{n=0}^{\infty}K(t_n). 
\end{equation}
By definition, $\Bar{A}_{\mu}(h)=\|A_\mu\|_{L^1(\mathbb{R}^+_0)}+\mathcal{O}(h)$ and $ \Bar{K}(h)=\|K\|_{L^1(\mathbb{R}^+_0)}+\mathcal{O}(h)$ (see also \cite[Lemma 3.3]{Beh_discr_etna}). To obtain a structure-preserving discretization which retains the asymptotic behaviour of the solution to \eqref{model_beh}, we set
\begin{equation}\label{eq: factor gamma}
    \gamma(h)=\frac{1}{h\sum_{n\ge0}K(t_n)},
\end{equation}
so that the normalization condition $\Bar{K}(h)=1=\|K\|_{L^1(\mathbb{R}^+_0)}$ is exactly satisfied. Furthermore, we define the set
\begin{equation}\label{pos inv discr}
     D(h)= \{ [x, y, z]^\mathsf{T}\in\mathbb{R}^3 : 0< x \leq S_{max},\; 0 \leq y \leq F_{max}(h),\; 0 \leq z \leq M_{max}(h) \},
\end{equation}
where the upper bounds are given by
\begin{equation}\label{limit discr}
\begin{aligned}
     S_{\max} = N, \quad 
     F_{\max}(h) = \lambda \Big(\|A\|_{L^1(\mathbb{R}^+_0)}+\Bar{A}_{\mu}(h)\Big)
     + N\|A_{\mu}^\prime\|_{L^1(\mathbb{R}^+_0)},\quad M_{\max}(h) = 2g(F_{\max}(h)).
\end{aligned}
\end{equation}
The following result holds.

\begin{Theorem}\label{tb1}
Suppose that $\displaystyle\lim_{\tau\to\infty}A(\tau)=0$.
Then the region $D(h)$ introduced in \eqref{pos inv discr} is a positively invariant region for the discrete formulation \eqref{mod_discreto_short}.
\end{Theorem}

The discrete model \eqref{mod_discreto_short} always admits the disease-free equilibrium $DFE(h)$ which corresponds to the extinction of the infection in the population. Moreover, whenever $R_0(h)>1,$ with
\begin{equation}\label{eq:Claudia_Roh}
    R_0(h)=N\,\bar{A}_{\mu}(h), 
\end{equation}
and independently of the choice of the kernel $K$, the system \eqref{mod_discreto_short} also admits an endemic equilibrium $EE(h)$ (cf. \cite[Theorem~4.1]{Beh_discr_etna}).  We remark that $R_0(h)=R_0+\mathcal{O}(h)$ in \eqref{eq:Claudia_Roh} represents a numerical approximation of the basic reproduction number $R_0.$ The stability analysis of the discrete equilibrium $EE(h)$ goes through the linearization theory in \cite{baker2003}. Additional details on the topic are provided in \cite{Beh_discr_etna}. In Section \ref{subsec: example beh}, numerical examples for selected kernels illustrate the asymptotic behaviour of the solutions of the discrete-time model.

\section{Tool description and core kernels}\label{sec:Tool_Description}
In this section, we present the software packages we developed for the numerical methods introduced in Section \ref{sec:Solutions and Algorithms}. The codes are freely available at \href{https://github.com/ghitan/EPITIME}{github.com/ghitan/EPITIME}. Our implementations are designed to prioritize vectorized operations and adopt a modular organization. They are structured into clearly identified code blocks (CB), which are consistently referenced throughout this section. Each block corresponds to a specific component of the package and is labelled \texttt{MCB:A} \ – \ \texttt{MCB:G} in MATLAB and \texttt{PCB:A} \ – \ \texttt{PCB:G} in Python.

\subsection{The NSFD\_AoI software}\label{subsec:AoI_Implementation}
The NSFD scheme \eqref{eq:NSFDscheme} is implemented in MATLAB and Python within the function \texttt{NSFD\_AoI}. Each solver accepts either a structured input (struct/dict) or individual arguments and automatically assigns default values to any missing fields. The main inputs of our routines are:
\begin{itemize}
\item the initial susceptible population $S_0$ and the total population size $N,$
\item the final time $T$ and the discretization step-size $h,$
\item the contact rate $\beta_0,$ the infectivity kernel $A(t)$ and the function $\varphi_0(t),$
\item a \texttt{verbosity} binary flag that controls warning messages (0 to suppress them, 1 to enable them).
\end{itemize}
The primary outputs are the discrete time vector \texttt{t} and the solution matrix \texttt{Y}, whose $n$-th column contains the discrete approximation $[S_n, \varphi_n]^\mathsf{T}$ at the time $t_n$. Upon request, the optional performance structure \texttt{P} is also returned. This object provides quantitative information on the execution. It is defined as a structured array in MATLAB and as a dictionary in Python, as detailed in Listings \ref{lst:MATP} and \ref{lst:PITP}, respectively.
\begin{lstlisting}[style=MatlabStyle, caption={MATLAB Optional performance matrix output (\texttt{MCB:E})}, label={lst:MATP}]
P = struct();
P.elapsed_time = elapsed_time;   % total CPU time (allocation and stepping)
P.steps_number = Nt - 1;         % total number of time steps
P.flops_number = flops_counter;  % estimated FLOPs
P.Kern_number  = Nt;             % number of kernel evaluations
P.memory_bytes = memory_bytes;   % estimated memory used by core arrays
\end{lstlisting}
\begin{lstlisting}[style=PythonStyle, caption={Python Optional performance matrix output (\texttt{PCB:E})}, label={lst:PITP}]
P = {
    'elapsed_time': elapsed_time,# total CPU time (allocation and stepping)
    'steps_number': Nt - 1,      # total number of time steps
    'flops_number': flops_count, # estimated FLOPs
    'Kern_number' : Nt,          # number of kernel evaluations
    'memory_bytes': memory_bytes # estimated memory used by core arrays
}
\end{lstlisting}
The total CPU time (\texttt{elapsed\_time}) is measured with the built-in \texttt{tic/toc} functions in MATLAB and with \texttt{time.time()} in Python. The memory usage (\texttt{memory\_bytes}) is estimated through the \texttt{whos} routine in MATLAB and through the \texttt{.nbytes} attribute in Python. The total number of floating-point operations (\texttt{flops\_number}) is computed by counting the arithmetic operations at each time step. This yields, at the $n$--th step, approximately $3n+5$ flops, which in turn produces a quadratic complexity in the number of temporal nodes, $\mathcal{O}(N_t^{2})$, for the full integration. The metric \texttt{Kern\_number} records the evaluations of the kernel function $A(t)$, which coincide with the size of the temporal grid. These quantities provide a basis for benchmarking the efficiency and resource usage of the implementation across different problem sizes.

The core time-stepping blocks implementing the NSFD update \eqref{eq:NSFDscheme} are shown in Listings \ref{lst:MATstep} and \ref{lst:Pystep}. Both versions rely on vectorized operations and compute the discrete convolution through optimized matrix-vector products, thus avoiding explicit summation loops. A minor implementation detail concerns the case in which the final time $T$ is not an integer multiple of the prescribed step-size $h.$ In this situation, the boolean flag \texttt{flag\_small} activates a final reduced step of size \texttt{h\_small}. This technical adjustment does not alter the convergence order of the NSFD method and does not affect its structure-preserving properties.
\begin{lstlisting}[style=MatlabStyle, caption={MATLAB time-stepping loop implementing the NSFD scheme (\texttt{MCB:D})}, label={lst:MATstep}]
tic;                     % H = AoI_problem.beta * h is previously assigned
for n = 1:Nt-1
    if flag_small == 1 && n == Nt-1
        H = AoI_problem.beta * h_small;
    end
    Y(1,n+1) = Y(1,n) / (1 + H * Y(2,n));              % NSFD update for S 
    tmp = (Y(2,1:n) .* Y(1,2:n+1)) * A_val(n+1:-1:2);  % Disc. convolution
    Y(2,n+1) = P0(n+1) + H * tmp;                      % NSFD update for (*@\textcolor{matlabgreen}{$\varphi$}@*)
end
stepping_time = toc;
elapsed_time = allocation_time + stepping_time;
\end{lstlisting}
\begin{lstlisting}[style=PythonStyle, caption={Python time-stepping loop implementing the NSFD scheme (\texttt{PCB:D})}, label={lst:Pystep}]
tic_step = time.time()               # H = beta * h is previously assigned
for n in range(Nt - 1):
    if flag_small and n == Nt - 2: H = beta * h_small
    Y[0, n + 1] = Y[0, n] / (1 + H * Y[1, n])          # NSFD update for S 
    conv = np.dot(Y[1, :n + 1] * Y[0, 1:n + 2], A_val[n + 1:0:-1, 0])
    Y[1, n + 1] = P0[n + 1] + H * conv                 # NSFD update for (*@\textcolor{yellow!50!black}{$\varphi$}@*)
step_time = time.time() - tic_step
elapsed_time = alloc_time + step_time
\end{lstlisting}
In addition to the core solver, the implementation relies on two auxiliary functions that handle input parsing and validation. Their purpose is to define default parameter values, complete missing entries and ensure the consistency of the problem definition before the time-stepping begins. The first auxiliary function, denoted \texttt{parse\_input} in MATLAB (\texttt{MCB:F}) and \texttt{\_parse\_input} in Python (\texttt{PCB:F}), collects the user-provided arguments and fills any missing values with predefined defaults. In MATLAB, the defaults are
\begin{lstlisting}[style=MatlabStyle, basicstyle=\ttfamily\small]
defaults = struct('S0', 990, 'N', 1000, 'T', 10, 'h', 0.1, 'beta', ...
                  2.5e-3, 'A', @(t) exp(-3*t), 'phi0', [], 'verbosity', 1);
\end{lstlisting}
and in Python the corresponding dictionary is 
\begin{lstlisting}[style=PythonStyle, basicstyle=\ttfamily\small]
defaults = {'S0': 990, 'N': 1000, 'T': 10, 'h': 0.1, 'beta': 2.5e-3,
            'A': lambda t: np.exp(-3.0*t), 'phi0': None, 'verbosity': 1}
\end{lstlisting}
If the field \texttt{phi0} is not provided, it is automatically defined as $\varphi_0(t) = (N - S_0) A(t)$. The function also emits warnings for missing fields when the verbosity flag is active.

The second auxiliary function, \texttt{check\_input} in MATLAB (\texttt{MCB:G}) and \texttt{\_check\_input} in Python (\texttt{PCB:G}), verifies all input fields in accordance with the properties summarized in Table \ref{tab:Mario}. Scalar parameters such as $S_0$, $N$, $T$, $h$ and $\beta_0$ are required to be positive, finite and of the correct type. In MATLAB this is enforced via \texttt{validateattributes}, for example
\begin{lstlisting}[style=MatlabStyle, basicstyle=\ttfamily\small]
validateattributes(S0, {'numeric'}, {'scalar','positive','finite'}, ...
                   mfilename, 'S0');
\end{lstlisting}
while in Python an analogous test is
\begin{lstlisting}[style=PythonStyle, basicstyle=\ttfamily\small]
if S0 <= 0 or not np.isfinite(S0):
    raise ValueError('S0 must be positive and finite.')
\end{lstlisting}
Furthermore, vector inputs corresponding to the kernel $A(t)$ and the function $\varphi_0(t)$ are checked for non-negativity and elementwise consistency, i.e., $\varphi_0(t_n) \le (N - S_0) A(t_n)$ for each $n$. In MATLAB, this is implemented via vectorized comparisons
\begin{lstlisting}[style=MatlabStyle, basicstyle=\ttfamily\small]
if any(P0(:) > (N - S0) * A_val(:) + eps)
    warning('Some phi0 values exceed theoretical upper bound');
end
\end{lstlisting}
and in Python using NumPy operations
\begin{lstlisting}[style=PythonStyle, basicstyle=\ttfamily\small]
Eps = np.finfo(np.float64).eps
if np.any(P0.flatten() > (N - S0) * A_val.flatten() + Eps):
    print('Warning: Some phi0 values exceed theoretical upper bound')
\end{lstlisting}
Any violation of essential constraints triggers an error that immediately stops execution, whereas minor inconsistencies generate warnings if the verbosity flag is active. This layered validation ensures that the solver operates on a fully defined and consistent problem before entering the time-stepping phase.

\subsection{The NSFD\_BEH software}\label{subsec:Implementation beh}
The NSFD scheme \eqref{mod_discreto_short} for the behavioural model is developed in both MATLAB and Python within the function \texttt{NSFD\_BEH}. Specifically, our codes implement the non-standard discretization of \eqref{model_beh}, expressed in terms of the standardized quantities 
\begin{equation}\label{eq:normalized}
    s_n\approx\frac{S(t_n)}{N}, \qquad f_n\approx\frac{F(t_n)}{N}, \qquad m_n\approx\frac{M(t_n)}{N},
\end{equation}
with normalization factor $N:=\lambda/\mu,$ corresponding to the total population size. 

The \texttt{NSFD\_BEH} solver supports both structured inputs and explicit argument passing, with default values automatically assigned whenever optional parameters are omitted.  The primary inputs are described below:
\begin{itemize}
    \item the final time $T$ and the step-size $h$,
    \item the initial susceptible population $S_0$,
    \item the total population size $N$ and the natural death rate $\mu
    $, 
    \item the infectivity function $A_{\mu}(t)$, the memory kernel $K(t)$, the inhibition function $\beta(M)$ and the message function $g(F)$,
    \item a verbosity binary flag that controls warning messages (0 to suppress them, 1 to enable them).
\end{itemize}
As main outputs we retain only the epidemiologically most relevant quantities, $s_n$ and $f_n$, leaving out the auxiliary memory variable $m_n$. Accordingly, the solver returns the discrete time vector \texttt{t} and the solution matrix \texttt{Y}, whose $n$-th column stores the discrete approximations $[s_n, f_n]^\mathsf{T}$ as defined in \eqref{eq:normalized}.

 The corresponding approximations for the functions in the original model \eqref{model_beh} may be retrieved by multiplying these normalized values by the constant $N$.   
Additionally, an optional performance object \texttt{P} can be returned upon request. This object contains numerical details about the computation and is implemented as a structured array in MATLAB and as a dictionary in Python, 
as shown in Listing \ref{lst:MATP_beh} and \ref{lst:PITP_beh}, respectively.

\begin{lstlisting}[style=MatlabStyle, caption={MATLAB Optional performance matrix output (\texttt{MCB:E})}, label={lst:MATP_beh}]
P = struct();
P.elapsed_time  = elapsed_time;     % total CPU time
P.steps_number  = Nt;               % total number of time steps
P.flops_number  = flops_counter;    % estimated FLOPs
P.Kern_number   = Nt+K_eval;        % number of kernel evaluations
P.memory_bytes  = memory_bytes;     % estimated memory used by core arrays
\end{lstlisting}
\begin{lstlisting}[style=PythonStyle, caption={Python Optional performance matrix output (\texttt{PCB:E})}, label={lst:PITP_beh}]
P = {                                  
    'elapsed_time': elapsed_time,   # total CPU time
    'steps_number': Nt,             # total number of time steps
    'flops_number': flops_counter,  # estimated FLOPs
    'Kern_number' : Nt+K_eval,      # number of kernel evaluations
    'memory_bytes': memory_bytes    # estimated memory used by core arrays
}
\end{lstlisting}

The execution time (\texttt{elapsed\_time}) is recorded using native timing utilities in the two environments, namely the \texttt{tic/toc} pair in MATLAB and the \texttt{time.time()} function in Python. The amount of memory required by the algorithm (\texttt{memory\_bytes}) is evaluated by relying on built-in inspection tools, specifically the \texttt{whos} command in MATLAB and the \texttt{.nbytes} property in Python.
The total number of floating-point operations (\texttt{flops\_number}) is estimated by explicitly counting the arithmetic operations performed at each time step. The update of $S_{n+1}$ involves a fixed number of scalar operations, and the evaluation of the history-dependent terms requires the computation of two discrete convolutions. Overall, this results in approximately $6n+9$ floating-point operations per time step. Summing over all time levels yields a total computational cost that scales quadratically with the number of temporal nodes, leading to an overall complexity of $\mathcal{O}(N_t^{2})$.
The metric \texttt{Kern\_number} represents how many times the kernel functions $A_{\mu}(t)$ and $K(t)$ are evaluated. In particular, the number of evaluations of $A_{\mu}(t)$ matches the number of time nodes, whereas the number of evaluations of $K(t)$ is $K_{\text{eval}}$ with $K_{\text{eval}} \leq N_t^2$, due to the presence of multiple evaluations of $K$ required in the computation of the weight $\gamma(h)$.

The core time-stepping routines for the present method (see Listings \ref{lst:MATstep_beh} and \ref{lst:Pystep_beh}) adhere to the same implementation principles outlined in the previous section. A minor yet relevant difference is the preliminary computation of the non-standard weight $\gamma(h)$, which appears in equation $\eqref{mod_discreto_short}_3$ and is computed prior to advancing the solution in time. Its analytical expression is provided in~\eqref{eq: factor gamma}. 
In the numerical implementation, the infinite series defining $\gamma(h)$ is truncated by considering a sufficiently large number of terms which is proportional to $N_t^2$.
The subsequent updates are then performed using vectorized operations, with the history-dependent terms evaluated through optimized matrix--vector products, thus avoiding explicit summation loops.
As before, when the prescribed step-size $h$ does not exactly match the final time $T$, a reduced final step of length \texttt{h\_last} is employed. This technical detail preserves both the convergence order and the structure-preserving features of the underlying NSFD scheme.

\begin{lstlisting}[
    style=MatlabStyle,
    caption={MATLAB time-stepping loop implementing the NSFD scheme (\texttt{MCB:D})},
    label={lst:MATstep_beh}
]
tic
Sum1   = sum(K_vec);
T_incr = T + h;
incr   = h * BEH_prob.K(T_incr);
it_max = 0;
K_eval = Nt;

while incr > 1e-11 && it_max < Nt^2
    Sum1   = Sum1 + incr;
    T_incr = T_incr + h;
    incr   = h * BEH_prob.K(T_incr);
    it_max = it_max + 1;
    K_eval = K_eval + 1;
end

gammah = 1 / (h * Sum1);
weight_definition_time = toc;

% ----- Time iteration -----
tic
for n = 1:Nt-1

    if (D_tilde ~= 0 && n == Nt-1)
        D = D_tilde;
    end

    betaM_val = BEH_prob.betaM(N * M(n));
    g_val     = BEH_prob.g(N * Y(2,n));

    Y(1,n+1) = (Y(1,n) + mu * D) ...
               / (1 + D * (mu + N * betaM_val * Y(2,n)));

    betaM_vec(n) = N * betaM_val;
    g_vec(n)     = g_val;

    var1 = Y(1,2:n+1) .* betaM_vec(1:n) .* Y(2,1:n);
    int1 = A_vec(n+1:-1:2) * var1';
    int2 = K_vec(n:-1:1)   * g_vec(1:n)';

    Y(2,n+1) = F0_vec(n+1) + D * int1;
    M(n+1) = M0_vec(n+1) + gammah * (1/N) * D * int2;
             
end

stepping_time = toc;

elapsed_time = allocation_time + weight_definition_time + stepping_time;
\end{lstlisting}

\begin{lstlisting}[style=PythonStyle, caption={Python time-stepping loop implementing the NSFD scheme (\texttt{PCB:D})}, label={lst:Pystep_beh}]
start_weight = time.time()
Sum1   = np.sum(K_vec)
T_incr = T + h
incr   = h * BEH_prob['K'](T_incr)
it_max = 0
K_eval = Nt

while incr > 1e-11 and it_max < Nt**2:
    Sum1   += incr
    T_incr += h
    incr    = h * BEH_prob['K'](T_incr)
    it_max += 1
    K_eval += 1
    
gammah = 1.0 / (h * Sum1)
weight_definition_time = time.time() - start_weight

# ----- Time iteration -----
start_step = time.time()
for n in range(0, Nt-1):

    if D_tilde != 0 and n == Nt-2:
        D = D_tilde
        
    betaM_val = BEH_prob['betaM'](N * M[n])
    g_val     = BEH_prob['g'](N * Y[1, n])
    
    Y[0, n+1] = (Y[0, n] + mu * D) / \
                 (1 + D * (mu + N * betaM_val * Y[1, n]))
    betaM_vec[n] = N * betaM_val
    g_vec[n]     = g_val
    if n >= 1:
        var1 = Y[0, 1:n+1] * betaM_vec[0:n] * Y[1, 0:n]
        int1 = np.dot(A_vec[n:0:-1], var1)
        int2 = np.dot(K_vec[n-1::-1], g_vec[0:n])
    else:
        int1 = 0.0
        int2 = 0.0
    Y[1, n+1] = F0_vec[n+1] + D * int1
    M[n+1] = M0_vec[n+1] + gammah * (1.0 / N) * D * int2
    
stepping_time = time.time() - start_step
elapsed_time = allocation_time + weight_definition_time + stepping_time
\end{lstlisting}

Similarly to the \texttt{NSFD\_AoI} function, the solver \texttt{NSFD\_BEH} is complemented by auxiliary routines devoted to input parsing and validation. These auxiliary modules assign default values, complete missing parameters and verify the consistency of the problem setup before the time-stepping phase. Since their functionality mirrors that already described in  Section \ref{subsec:AoI_Implementation}, only the differences in the prescribed default values are reported here. 
The input-handling function is denoted by \texttt{parse\_input} in MATLAB (\texttt{MCB:F}) and the default parameters are set as follows:

\begin{lstlisting}[style=MatlabStyle, basicstyle=\ttfamily\small]
defaults = struct('T',1000,'h',1,'N',5e7,'mu',1/(75*365), ...
                  'S0',0.2*5e7,'A',[],'K',[],'betaM',[], ...
                  'g',[],'verbosity',1);
\end{lstlisting}
The corresponding function in Python is denoted as \texttt{\_parse\_input} (\texttt{PCB:F}), and the default parameters are
\begin{lstlisting}[style=PythonStyle, basicstyle=\ttfamily\small]
defaults = {'T': 1000, 'h': 1, 'N': 5e7, 'mu': 1 / (75 * 365),
            'S0': 0.2 * 5e7, 'A': None, 'K': None, 
            'betaM': None, 'g': None, 'verbosity': 1}
\end{lstlisting}

If any of the core functions are not provided, they are automatically initialized with the following default definitions
\begin{equation*}
    g(x)=x, \qquad A(t) = R_0 (\mu + \nu)^2 \, t \, e^{-(\mu+\nu)t}, \qquad K(t) = a \, e^{-a t}, \qquad \beta(x) = (1 + \alpha  x)^{-1}, {\color{teal}}
\end{equation*}
with $R_0=20$,  $\nu=1/7$,  $a=1/30$,  and $\alpha=8\cdot 10^3.$ When the verbosity flag is enabled, the routine also issues warnings for any missing optional fields. Furthermore, an auxiliary routine is employed to verify the validity of all input fields. In MATLAB, this is performed by the local function \texttt{check\_input} (\texttt{MCB:G}), 
and in Python by \texttt{\_check\_input} (\texttt{PCB:G}). 
The checks ensure that scalar parameters such as $S_0$, $N$, $T$, $h$ and 
$\mu$ are positive and finite,  using the same mechanisms described earlier (e.g., \texttt{validateattributes} in MATLAB).

\subsection{Software GUI}\label{subsec:GUI}
\begin{figure}[t]
  \centering
  \includegraphics[width=\linewidth]{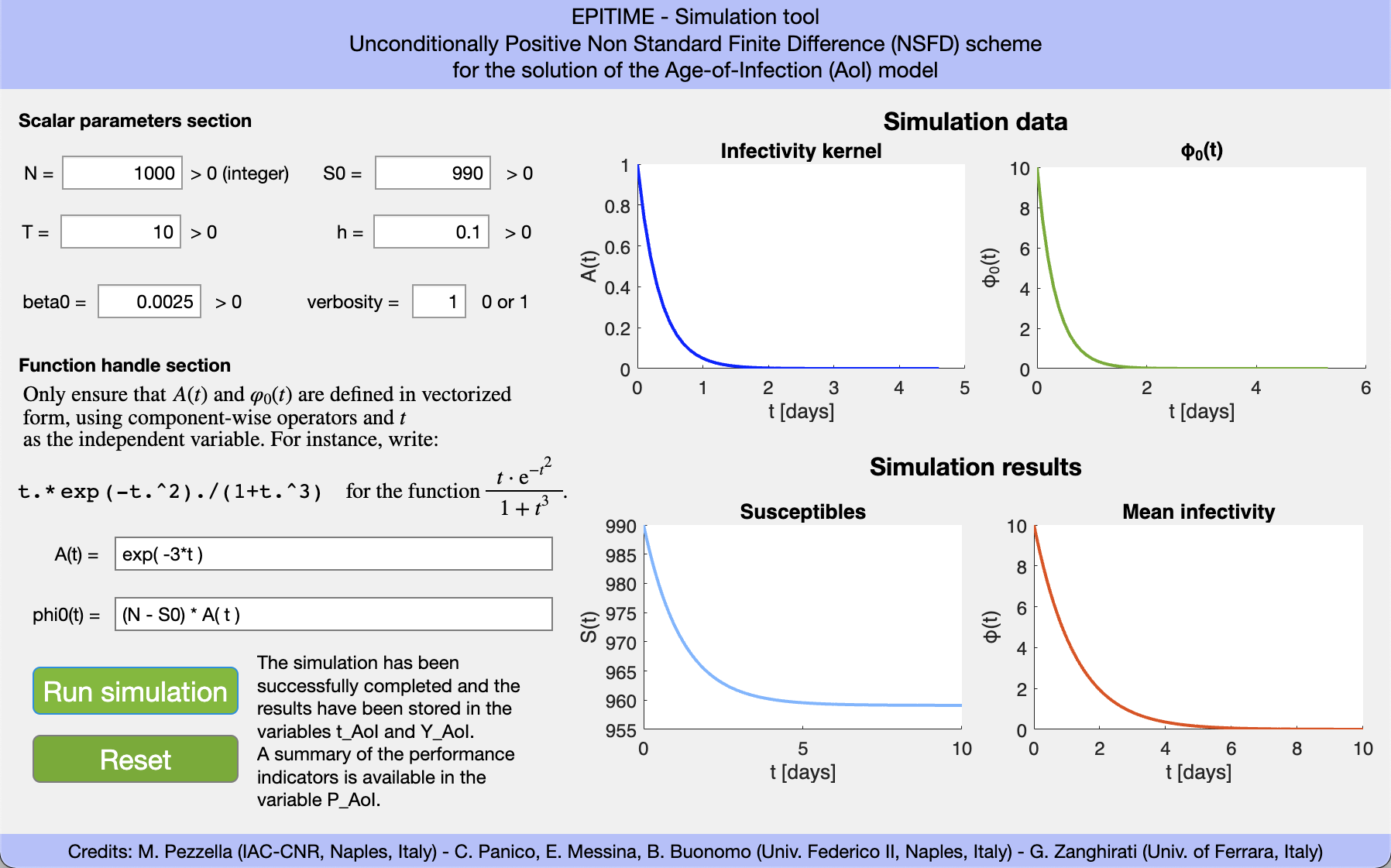}
  \caption{the software GUI of the AoI simulation tool after a successful run with the default  model functions  $A(t)=\exp(-3t)$ and $\phi_0(t)=(N-S_0)A(t),$ and parameters $N=1000$, $S_0=990$, $T=10$, $h=0.1$, and $\beta_0=0.0025.$
  }
  \label{fig:GUI}
\end{figure}
This section briefly introduces the MATLAB Graphical User Interface (GUI) developed for the proposed software. Its purpose is to facilitate simulations for users who are not particularly familiar with MATLAB or Python programming. The GUI can be launched from the \emph{Apps} panel in the MATLAB main window by opening the \texttt{SimulationTool.mlapp} file. Alternatively, it can be opened from the standard file browser by navigating to the folder containing the \texttt{.mlapp} file and all the associated resources.

The GUI is organized into two main panels (Figure \ref{fig:GUI}): 
the left panel accepts the problem inputs, while the right panel 
contains the plots. A header and a footer panel
complete the GUI: the former is for the software title and subtitle,
the latter is for the credits.

The upper part of the left panel is devoted to the parameter 
settings. The input parameters are the same as
those required by the already described main M-function 
\texttt{NSFD\_AoI}, as detailed in Section~\ref{subsec:AoI_Implementation}. Each numeric input field is initialized with its
default value. The user is allowed to change those values,
constrained to the corresponding type and range.

The central part of the left panel contains two text fields for the given functions $A(t)$ and $\varphi_{0}(t)$ in \eqref{eq:AoI_Infectivity} (see Table~\ref{tab:Mario} for their definitions and further details). These input fields accept string input corresponding to the Matlab code implementing the two model functions described above. The provided expressions are automatically converted into function handles, which are then passed to the \texttt{NSFD\_AoI} main simulation function. A small example is shown as an accompanying text to help the user. The easiest way to use these two fields is to directly type in  the Matlab code, which is the usual case for simple, one-line functions. The relevant requirements are that the function variable is denoted by \texttt{t} and that both functions are coded in a vectorized way, as illustrated in the accompanying text example. Although both inputs fields can be stretched by switching the GUI to full-screen mode to allow very long lines of code, this is not recommended. Alternatively, for more flexible or more complex expressions, the user may implement the code in external MATLAB function files and provide the corresponding function calls in the appropriate GUI input field. Even in this case, it is expected that the user-provided code returns a vector output, if the M-function is called with a vector input \texttt{t}. 
The function call will always be internally converted
into a function handle.
The default settings for the two model 
functions are  
$A(t)=\exp(-3t)$ and $\phi_0(t)=(N-S_0)A(t),$ with parameters $N=1000$, $S_0=990$, $T=10$, $h=0.1$, and $\beta_0=0.0025$
(see Section~\ref{subsec:AoI_Implementation}).
With these settings, the graphical output generated by the simulation is shown in Figure \ref{fig:GUI}.

The lower part of the left panel contains the simulation controls. A \texttt{Run simulation} button allows the simulation to be started, while a \texttt{Reset} button can be used to restore the input fields to their default values. To the right of the two buttons, a text area displays messages produced by the main \texttt{NSFD\_AoI} M-function at the end of the simulation, as well as any error or warning messages.

The right panel of the GUI contains the relevant plots. The upper part shows the profiles of the two model functions $A(t)$ (left) and $\varphi_{0}(t)$ (right). These plots are updated whenever the corresponding expressions are entered in the respective text fields. The lower part of the right panel displays the simulation results, namely the susceptibility $S(t)$ and the mean infectivity $\phi(t)$ up to time $T$. These plots are updated only at the end of the simulation and correspond to the two rows of the output array \texttt{Y} returned by the \texttt{NSFD\_AoI} M-function. The full \texttt{NSFD\_AoI} output is automatically made available in the MATLAB main workspace under the names
\texttt{t\_AoI} (time steps), \texttt{Y\_AoI} (solution values), and \texttt{P\_AoI} (performance data).

\begin{figure}[t]
  \centering
  \includegraphics[width=\linewidth]{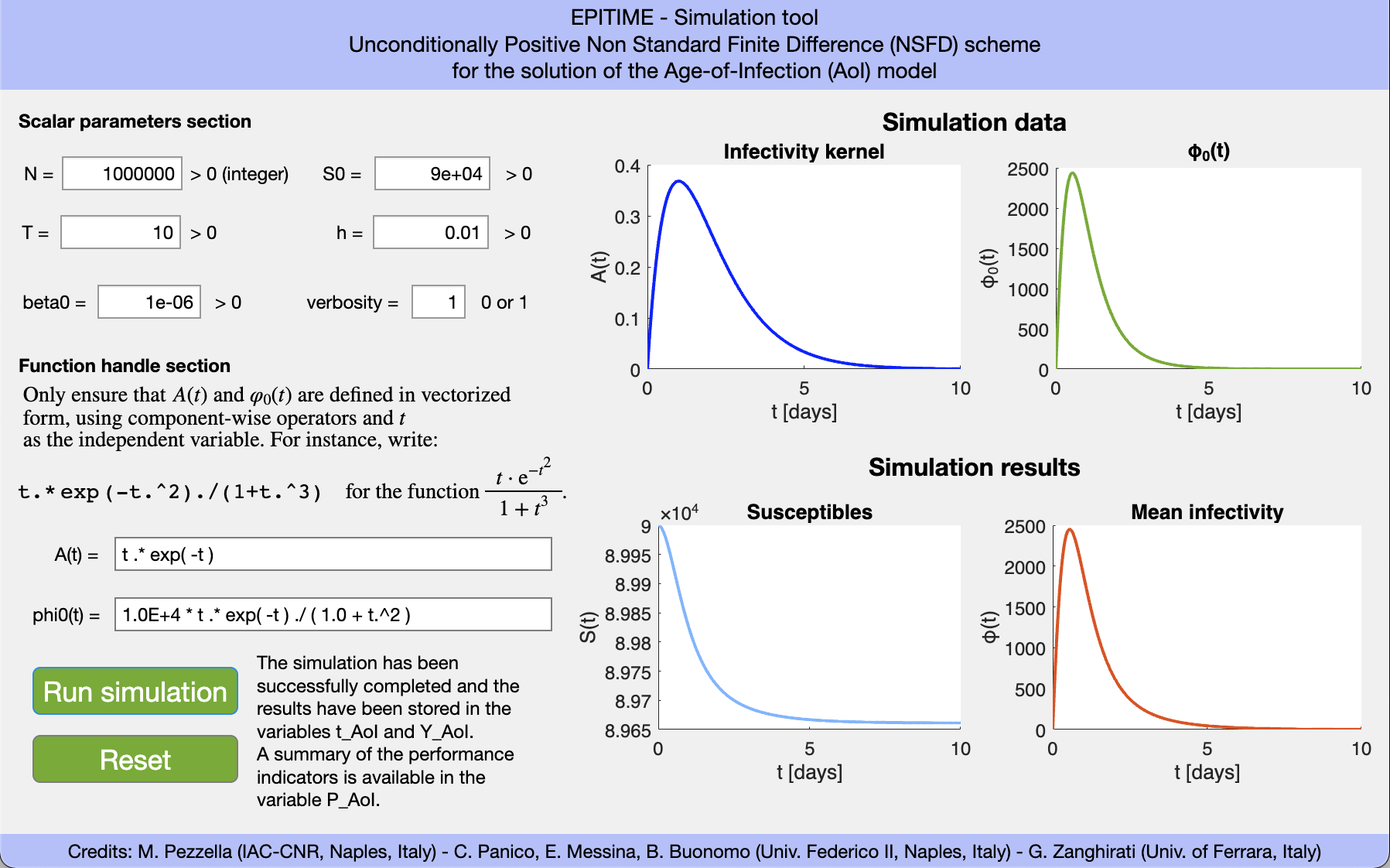}
\caption{the software GUI of the AoI simulation tool after a successful run with scalar parameters $N=10^6$, $S_0=9\cdot 10^4$, $T=10$, $h=10^{-2}$, $\beta_0=10^{-6}$. The model functions are set to $A(t)=t e^{-t}$ and $\phi_0(t)=10^4 t e^{-t}/(1+t^2)$.} 
  \label{fig:GUI1}
\end{figure}

\begin{figure}[t]
  \centering
  \includegraphics[width=\linewidth]{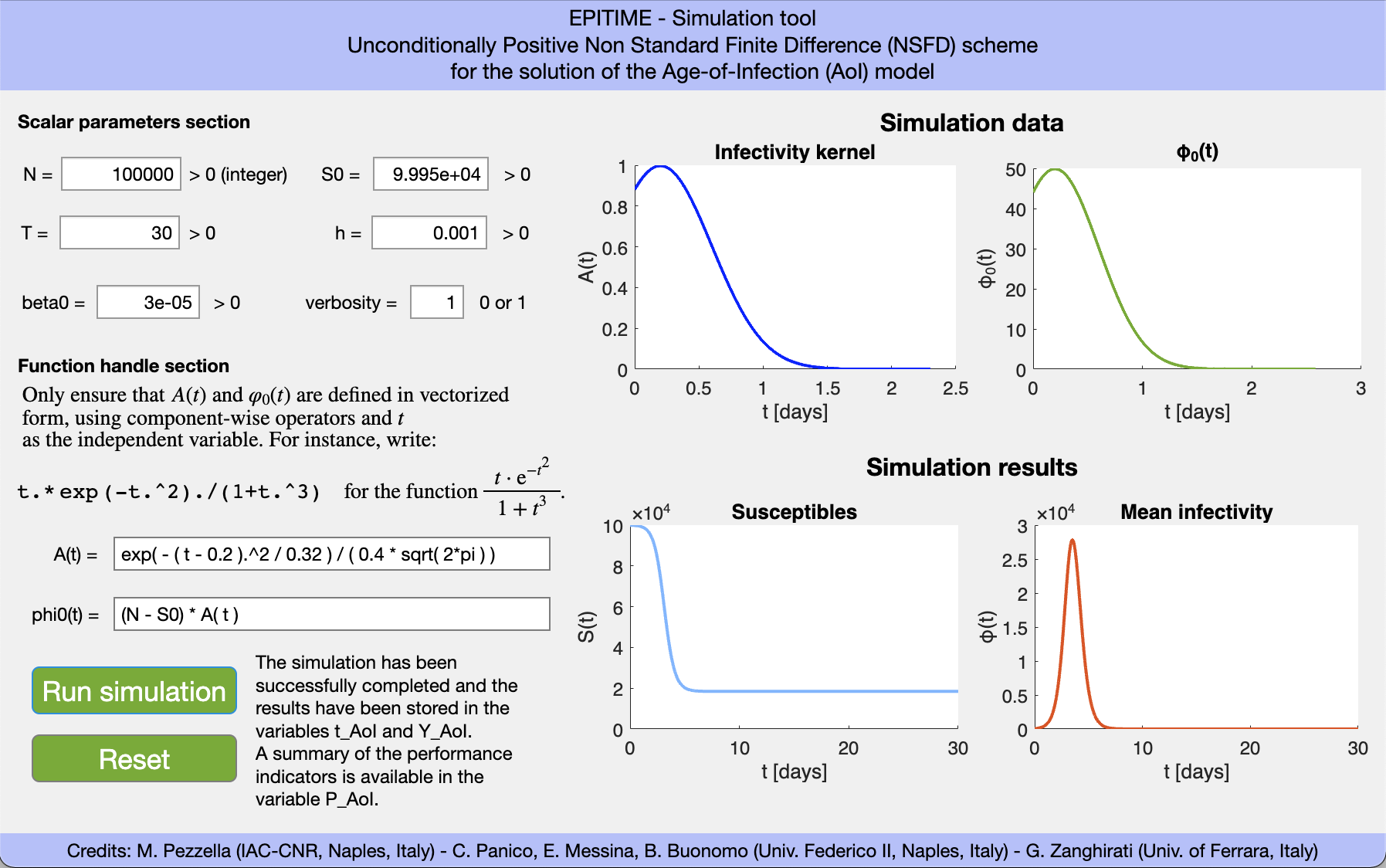}
  \caption{the software GUI of the AoI simulation tool after a successful run corresponding to the test case in \eqref{eq:Test2} of Section \ref{subsec:AoI_Asympt}. }
\label{fig:GUI2}
\end{figure}

To illustrate the use of the proposed GUI, two representative test cases are presented. Figure \ref{fig:GUI1} shows the simulation output obtained with $A(t)=t e^{-t}$ and $\phi_0(t)=10^4 t e^{-t}/(1+t^2)$, for $N=10^6$, $S_0=9\cdot 10^4$, $T=10$, $h=10^{-2}$, and $\beta_0=10^{-6}$. Figure \ref{fig:GUI2} displays the output corresponding to the test case in \eqref{eq:Test2} of Section \ref{subsec:AoI_Asympt}.

As a complementary development, a lightweight Python implementation of the GUI is shared in the \href{https://github.com/ghitan/EPITIME}{GitHub repository} as an interactive Jupyter notebook (\texttt{AoI\_Simulation\_Tool.ipynb}), with functionalities equivalent to those of the MATLAB interface.

\section{Numerical experiments and case studies}\label{sec:Numerical_Experiments}
In this section, we present numerical experiments designed to illustrate the accuracy, structure-preserving properties, and practical applicability of the proposed schemes. 
\subsection{The final size and the basic reproduction number}\label{subsec:AoI_Asympt}
Our first test case examines the asymptotic behaviour of the NSFD solution to \eqref{eq:AoI_Infectivity}, with particular emphasis on the convergence of the numerical final size towards its continuous counterpart. To begin with, we recall that the \emph{final size} of the epidemic, defined as $S(\infty) = \lim\limits_{t \to +\infty} S(t)$, satisfies the non-linear relation
\begin{equation}\label{eq:Final_size_scalare}
    \log\left(\frac{S_0}{S(\infty)}\right) = \left(1-\frac{S(\infty)}{N}\right) R_0 + \beta_0 \int_0^{\infty} \big(\varphi_0(\tau)-(N-S_0) A(\tau)\big) \, d\tau,
\end{equation}
where 
\begin{equation}\label{eq:R_0}
    R_0 = \beta_0 N \int_0^{\infty} A(\tau) \, d\tau
\end{equation}
is the \emph{basic reproduction number}, that is the expected number of secondary cases generated by a single primary case in a fully susceptible population in the absence of interventions \cite{BrauerBook}. Furthermore, under suitable regularity assumptions (see, for instance, \cite[Section 3]{Brauer2005}) it follows that $\varphi(t) \to 0$ as $t \to \infty$.

\begin{figure}[!t]
\centering
\includegraphics[scale=0.70]{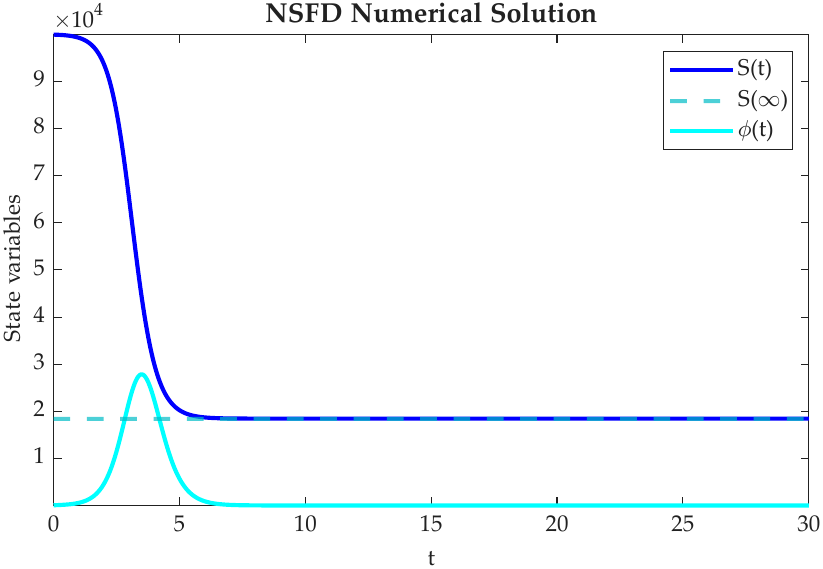}
\caption{NSFD numerical solution of problem \eqref{eq:AoI_Infectivity}--\eqref{eq:Test2} for $h=10^{-3}$.}
\label{fig:Test2_Paper_JCD}
\end{figure}

At the discrete level, the \emph{numerical final size} $S_\infty(h)$ is guaranteed to exist by Theorem \ref{THMmimimcNSFD}. In the present test, the asymptotic behaviour of the NSFD solution is approximated by the value $\tilde{S}_\infty(h)$ of the numerical simulation at a sufficiently large final time $T$. The corresponding \emph{numerical basic reproduction number} is defined as
\begin{equation*}
    R_0(h) = h \beta_0 N \sum_{n=0}^{\infty} A(t_{n+1}). 
\end{equation*}
We note that, in addition to the assumptions listed in Table~\ref{tab:Mario}, if
\begin{equation*}
    A^{\prime}\in L^1(\mathbb{R}^+_0)
    \qquad \text{or} \qquad
    \exists\,\mathcal{T}\in\mathbb{R}^+ \ \text{such that} \ A(t)\ \text{is monotonic for all}\ t\geq\mathcal{T},
\end{equation*}
then $R_0(h)$ converges linearly to $R_0$ as the step-size tends to zero
(cf. \cite[Lemma 3.3]{NSFD_MPV} and \cite[p. 208]{Davis}). Furthermore, it has been shown in \cite{NSFD_MPV} that $R_0(h)$ plays, for the discrete dynamics, a role analogous to that of $R_0$ in the continuous model, and that a discrete counterpart of the classical final-size relation \eqref{eq:Final_size_scalare} holds \cite[Theorem 4.4]{NSFD_MPV}. To illustrate these theoretical considerations, we consider the problem defined by
\begin{equation}\label{eq:Test2}
    N = 10^5, \hfill S_0 = 99950, \hfill \beta_0 = 3 \cdot 10^{-5}, \hfill T = 30, \hfill A(t) = \frac{1}{0.4 \sqrt{2\pi}} \exp\left(-\frac{(t-0.2)^2}{2\cdot0.4^2}\right).
\end{equation}
Specifically, we address the scenario in which $\varphi_0(t)$ attains its upper bound in \eqref{eq:Bound_phi0} for all $t.$ As a result, the values $S(\infty) = 1.8389 \cdot 10^4$ and $R_0=2.0744$ follow from \eqref{eq:Final_size_scalare} and \eqref{eq:R_0}, respectively. The numerical solution computed by the NSFD integrator \eqref{eq:NSFDscheme} with step-size $h = 10^{-3}$ is displayed in Figure~\ref{fig:Test2_Paper_JCD}.
The NSFD scheme \eqref{eq:NSFDscheme} is applied for four progressively refined step-sizes $h = 10^{-i},$ with $i=1,\dots,4$, and the values of $R_0(h),$ $\tilde{S}_\infty(h)$, and $\varphi_\infty(h)$ are computed for each discretization level. The simulation results, reported in Table \ref{tab:Ex1}, show that the numerical final size $\tilde{S}_\infty(h)$ converges linearly to $S(\infty)$ as the step-size decreases. As outlined in \cite{ImplicitVIE}, the convergence of the asymptotic limit is a non-trivial property arising from the dynamical consistency of the method (see also \cite[Theorem 4.1]{NSFD_MPV}). Additionally, $\varphi_\infty(h)$ attains values close to machine precision, confirming the expected extinction of the total infectivity. 

\begin{table}[H] 
\caption{Numerical basic reproduction number and NSFD solution asymptotic behaviour for different step-sizes $h$.}
\label{tab:Ex1}
\newcolumntype{C}{>{\centering\arraybackslash}X}
\begin{tabularx}{\textwidth}{CCCCC}
\toprule
\textbf{$h$} & \textbf{$R_0(h)$} & \textbf{$\tilde{S}_\infty(h)$} & Rel. S$_\infty$ Err. & \textbf{$\varphi_\infty(h)$} \\
\midrule
$10^{-1}$ & $1.93960$ & $2.32114 \cdot 10^4$ & $2.62268 \cdot 10^{-1}$ & $7.03736 \cdot 10^{-14}$ \\
$10^{-2}$ & $2.06116$ & $1.88521 \cdot 10^4$ & $2.52009 \cdot 10^{-2}$ & $9.72951 \cdot 10^{-18}$ \\
$10^{-3}$ & $2.07307$ & $1.84349 \cdot 10^4$ & $2.51243 \cdot 10^{-3}$ & $3.67883 \cdot 10^{-18}$ \\
$10^{-4}$ & $2.07426$ & $1.83933 \cdot 10^4$ & $2.51169 \cdot 10^{-4}$ & $3.33460 \cdot 10^{-18}$ \\
\bottomrule
\end{tabularx}
\end{table}

\subsection{An Inverse Problem Framework for Inferring COVID-19 Infectivity Kernel}

Despite the modeling potential of the integro-differential system \eqref{eq:AoI_Infectivity}, the identification of a contact rate $\beta_0$ and an infectivity kernel $A$ that accurately reflect a real epidemic scenario remains a challenging task. Similar problems have been addressed in different settings, including the reconstruction of time-varying transmission rates \cite{Pijpers_2021}, inverse formulations for integral epidemic models \cite{Hritonenko_2022} and data-driven identification of distributed memory kernels \cite{Breda2025}. In this section, we propose a case study to investigate the reliability of the EPITIME software as a tool to support the reconstruction of infectivity-kernel profiles in a real-world epidemic setting. To overcome this challenge, we introduce a functional optimization procedure informed by empirical incidence and demographic data. In particular, we consider the daily number of newly reported COVID-19 cases across Italy, as reported by \emph{Il Sole 24 Ore} \cite{IlSole24Ore2025}, covering the period from 24 February 2020 to 8 January 2025 ($T = 1781$ days), along with the total resident population $N = 58934177$, according to the demographic data provided by Eurostat \cite{EurostatPopulation2025}.

\begin{figure}[!t]
\centering
\includegraphics[scale=0.70]{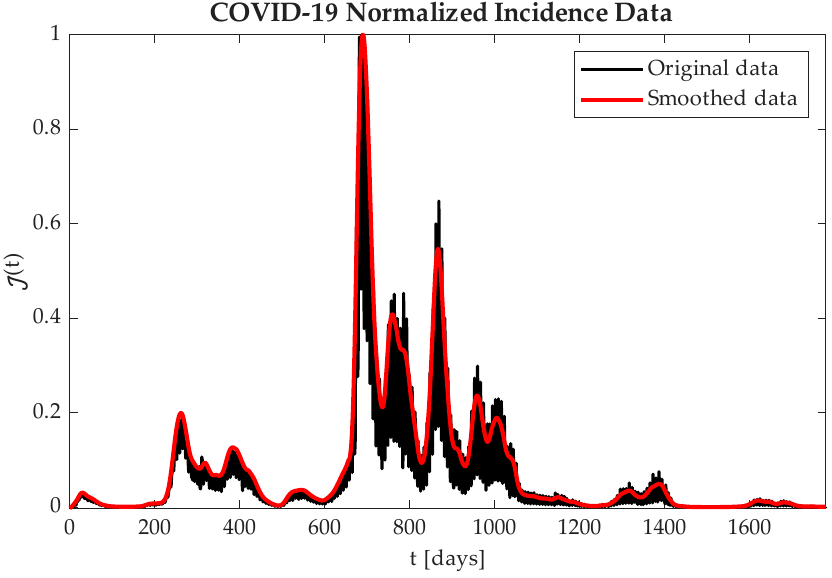}
\caption{Normalized daily COVID-19 incidence in Italy (24 February 2020 – 8 January 2025), including raw data from \cite{IlSole24Ore2025} and the corresponding smoothed profile obtained via quadratic regression. Here, the normalization is performed with respect to the maximum value, yielding $\kappa_{\mathcal{J}} \approx 5.68 \cdot 10^{-6}$.}
\label{fig:Covid_Data}
\end{figure}

Focusing on the first equation of model \eqref{eq:AoI_Infectivity}, the normalized incidence and its NSFD approximation are defined as
\begin{equation*}
\mathcal{J}(t) = \kappa_{\mathcal{J}}\beta_0 S(t)\varphi(t), \quad t \in [0,T],
\qquad \text{and} \qquad
\mathcal{J}_n = \kappa_{\mathcal{J}}\beta_0 S_n\varphi_n, \quad n = 0,\dots,N_t,
\end{equation*}
where $\kappa_{\mathcal{J}}$ is a normalization constant extracted from the data. Let $\mathcal{J}_i^{\mathrm{data}} \approx \mathcal{J}(\tau_i)$, for $i = 0, \dots, N_{\mathrm{data}}$, denote the empirical incidence at time $\tau_i$, obtained from \cite{IlSole24Ore2025} through quadratic regression smoothing followed by normalization (see Figure \ref{fig:Covid_Data}). These empirical values are compared with the simulated sequence $\{\mathcal{J}_{n_i}\}_{i=0}^{N_{\mathrm{data}}}$, where $n_i$ satisfies $\tau_i = n_i h$.

To select a functional form for the kernel $A(t)$, we expand it in terms of shifted Bernstein basis polynomials (see, e.g., \cite{Lorentz1986-zy}). Given a positive $M\in\mathbb{N},$ representing the maximum polynomial degree, we assume
\begin{equation*}\label{eq:Bernstein-Kernel}
    A(t) = \! \sum_{m=0}^{M} \kappa_m \, B_{m}^M\!\left(\frac{t}{T}\right)
    = \! \sum_{m=0}^{M} \kappa_m \, \binom{M}{m}
    \left(\frac{t}{T}\right)^{\! m} \!
    \left(1 - \frac{t}{T}\right)^{\! M-m}\!\!\!,
    \quad \varphi_0(t)=(N-S_0)A(t), \quad t \in [0,T],
\end{equation*}
where the coefficients $\kappa_m$, $m = 0,\dots,M$, are unknown parameters to be determined. Owing to the non-negativity of the Bernstein basis, $A(t)$ is guaranteed to remain non-negative if $\kappa_m \geq 0$ for all $m$. Consequently, we formulate a calibration problem for the parameter vector $\bm{p}=[\beta_0,\kappa_0,\dots,\kappa_M]\in\mathbb{R}_+^{M+2}$, aiming to minimize the discrepancy between observed and simulated incidence data. The optimal parameter set $\bm{p}^*$ is defined as the solution to the constrained optimization problem
\begin{equation}\label{eq:optimization}
    \begin{aligned}
    & \min_{\bm{p} \in \mathbb{R}_+^{M+2}}  
    \left( w_{L^2} \sum_{i=0}^{N_{\mathrm{data}}} 
    \dfrac{\bigl(\mathcal{J}_i^{\mathrm{data}} - \mathcal{J}_{n_i}(\bm{p})\bigr)}{N_{\mathrm{data}}+1}^2 
    + w_{\mathrm{peak}} \bigl(\mathcal{J}_{i_{\mathrm{peak}}}^{\mathrm{data}} - \mathcal{J}_{n_{i_{\mathrm{peak}}}}(\bm{p})\bigr)^2 + w_{\mathrm{initial}} \bigl(\mathcal{J}_0^{\mathrm{data}} - \mathcal{J}_{n_0}(\bm{p})\bigr)^2 \right. \\
    &  \left. \phantom{\min_{\bm{p} \in \mathbb{R}_+^{M+2}}   } \ \ 
    + w_{\mathrm{year}} \sum_{i=0}^{364} \dfrac{\bigl(\mathcal{J}_i^{\mathrm{data}} - \mathcal{J}_{n_i}(\bm{p})\bigr)^2}{365} \right)
    \\
    & \ \ \textrm{subject to} \   \tilde{R}_0(\bm{p}) = \frac{\beta_0 N T}{M+1} \sum_{m=0}^{M} \kappa_m \in \left[R_0^{\textrm{lower}}, R_0^\textrm{upper}\right],
    \end{aligned}
\end{equation}
which yields the data-informed contact rate $\beta_0^*$ and infectivity kernel $A^*(t) = \sum_{m=0}^{M} \kappa_m^* \, B_{m}^M(t/T).$ The objective functional in \eqref{eq:optimization} combines four terms, each emphasizing a distinct aspect of the epidemic dynamics
\begin{itemize}
\item a mean squared error over the entire dataset, measuring the global discrepancy between observed and simulated incidences;
\item a peak error term, enforcing accuracy at the time of maximum observed incidence;
\item an initial-time penalty, ensuring consistency at the onset of the epidemic;
\item a yearly error term, enforcing accurate reconstruction of early-stage dynamics over the first 365 days.
\end{itemize}
The weights $w_{L^2}$, $w_{\mathrm{peak}}$, $w_{\mathrm{initial}}$ and $w_{\mathrm{year}}$ are determined according to a reciprocal-optimum strategy. Each weight is set as the reciprocal of the minimum value attained by the corresponding objective component when optimized independently (all other weights set to zero), ensuring balanced contributions within the overall functional. The additional constraint enforces positivity of the coefficients and restricts the truncated basic reproduction number
\begin{equation*}
    \tilde{R}_0 =\beta_0 N\int_0^TA(t)\ dt = \dfrac{\beta_0 N T}{M+1} \sum_{m=0}^{M} \kappa_m,
\end{equation*}
to lie within a prescribed plausible range. These bounds, set to $R_0^{\textrm{lower}}=5 \cdot 10^{-10}$ and $R_0^{\textrm{upper}}=5 \cdot 10^1$ in our case, may be refined if more precise estimates of the basic reproduction number become available. 

\begin{figure}[!t]
\centering
\includegraphics[scale=0.9]{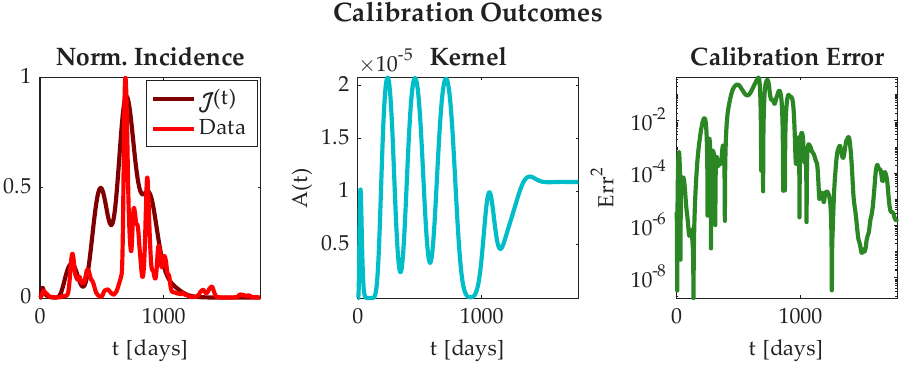}
\caption{Comparison of the empirical COVID-19 incidence in Italy with the simulated incidence (left panel), the corresponding estimated infectivity kernel (middle panel) and the pointwise quadratic error relative to the data (right panel). Here, the optimized contact rate is $\beta_0^* = 2.18 \cdot 10^{-5}$.}
\label{fig:Ar_Cal_Outcomes}
\end{figure}

The model calibration problem formulated in \eqref{eq:optimization} is addressed in Section 3.4 of the MATLAB live script \texttt{NSFD\_AoI\_LIVE}. This implementation allows the user to select both the dimension of the Bernstein polynomial space (we take $M=500$) and the optimization algorithm. Notably, the optimization procedure requires repeated evaluations of the direct temporal integration problem, making computational efficiency and long-term accuracy indispensable. These requirements are naturally met by the NSFD scheme \eqref{eq:NSFDscheme}, which provides stable and reliable results even over extended time horizons. For the present study, the optimization is performed using MATLAB’s \texttt{fmincon} routine with \texttt{StepTolerance} and \texttt{OptimalityTolerance} set to $10^{-14}$, \texttt{ConstraintTolerance} equal to $10^{-16}$ and automatic differentiation activated wherever possible to enhance efficiency. Furthermore, the solver \texttt{NSFD\_AoI} is run with a time step $h=0.25,$ which is smaller than the minimum interval between consecutive measurements. The results of the calibration are summarized in Figure \ref{fig:Ar_Cal_Outcomes}, where the simulated incidence is compared with the empirical data and the estimated infectivity kernel is depicted. The figure also reports the pointwise quadratic absolute error, whose mean value of $4.2\%$ supports the robustness and accuracy of the implemented methodology in reconstructing the dynamics of a real epidemic.

\subsection{Qualitative properties of solutions to the integral behavioural epidemic model}\label{subsec: example beh}
Before presenting the numerical experiments, we briefly recall the main qualitative properties of the behavioural integral epidemic model \eqref{model_beh} and of its  discrete counterpart \eqref{mod_discreto_short}. Under the assumptions stated in Section~\ref{subsec:NSFD_BEH}, the continuous model admits a positively invariant region, always possesses the disease-free equilibrium DFE, and admits a unique endemic equilibrium EE whenever $R_0>1$. Moreover, the local stability of the equilibria depends on the specific choice of the infectivity function $A_\mu$ and the memory kernel $K$. Specifically, suitable classes of kernels guarantee local asymptotic stability of the endemic state, whereas more structured memory effects may lead to instability and sustained oscillations \cite{BehRE_jmb}. 
The non-standard discretization~\eqref{mod_discreto_short} is designed so as to preserve these structural features: in particular, it maintains positivity and boundedness for every $h>0$, always admits a disease-free equilibrium, and admits a unique endemic equilibrium whenever the discrete threshold condition $R_0(h)>1$ is satisfied, with $R_0(h)$ providing a first-order approximation of $R_0.$ In addition, the discrete characteristic equation is consistent with the continuous one as $h\to0$, and the stability conditions derived for the numerical equilibria converge to the corresponding continuous conditions (see the results in~\cite{Beh_discr_etna} for a detailed analysis). The simulations reported below are intended to illustrate precisely this qualitative coherence between the continuous behavioural model and its NSFD approximation. 
\\
In this section, we illustrate two numerical examples based on the non-standard discretization~\eqref{mod_discreto_short} of model~\eqref{model_beh}. For all simulations, we adopt
\begin{equation*}
    g(F)=F, \qquad
    \beta(M)=\frac{1}{1+\alpha M}.
\end{equation*}
The infectivity function $A_\mu(t)$ and the memory kernel $K(t)$ are specified according to the particular scenario under consideration. The first experiment illustrates convergence toward the endemic equilibrium in a regime where stability is theoretically expected, whereas the second one highlights the onset of oscillatory behaviour induced by a more structured memory kernel.
\\ 
For the first numerical experiment, we consider the following parameters:
\begin{equation}\label{eq:Test1 beh}
    T=1000, \qquad N=5\cdot 10^{7}, \qquad
    \mu=\frac{1}{75\cdot 365}\,\mathrm{days}^{-1}, \qquad
    S_0=0.99\,S_e, \qquad \alpha=8\cdot 10^{3},
\end{equation}
where $S_e$ denotes the first coordinate of the endemic equilibrium. Furthermore, we assume that the infectivity function exhibits a unimodal profile and is given by
\begin{equation}\label{eq:test1 A beh}
    A(t) = \beta_0 t e^{-\nu t},
\end{equation}
with $ A_{\mu}(t) = e^{-\mu t}A(t),$ $\nu = 1/7\text{ days}^{-1}.$ We set $R_0 = 20$, and the parameter $\beta_0$ is determined so as to satisfy the condition for the basic reproduction number reported in Table \ref{tab:Claudia}, that is $\beta_0=(R_0/N)(\mu+\nu)^2$.

\begin{figure}[!t]
\centering
\includegraphics[scale=0.70]{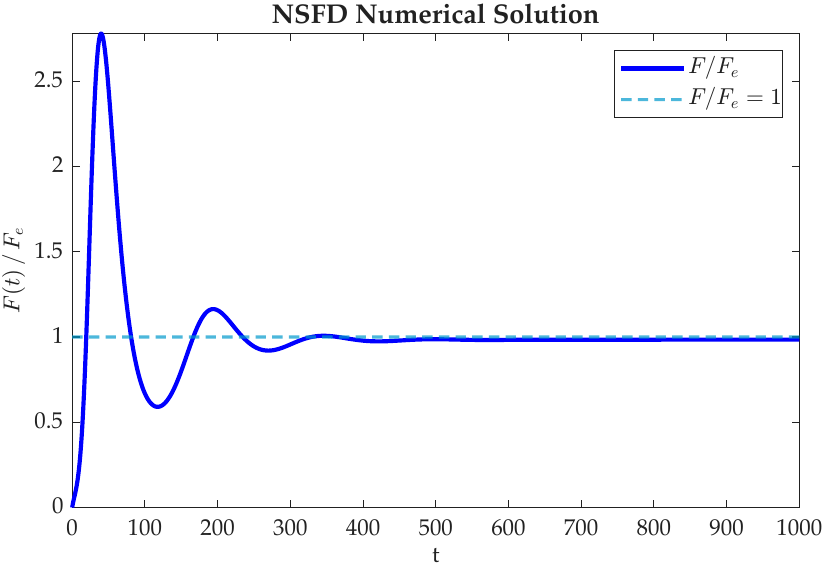}
\caption{NSFD numerical solution of problem \eqref{mod_discreto_short}--\eqref{eq:Test1 beh} for $h=10^{-1}$. The infectivity function and the memory kernel are set as detailed in \eqref{eq:test1 A beh}  and \eqref{eq: memory kernel n1}, respectively.}
\label{fig:Test1_Paper_JMB_beh}
\end{figure}

In Figure~\ref{fig:Test1_Paper_JMB_beh}, we present the time evolution of the driving factor of the FoI, $F(t)$, normalized with respect to its endemic equilibrium value $F_e$. The results are obtained using the parameters in~\eqref{eq:Test1 beh}, the infectivity function defined in~\eqref{eq:test1 A beh}, and the memory kernel
\begin{equation}\label{eq: memory kernel n1}
    K(t) = a e^{-a t},
\end{equation}
with $a=1/30\text{ days}^{-1}$. As shown in Figure~\ref{fig:Test1_Paper_JMB_beh}, the ratio $F(t)/F_e$ converges asymptotically to one, indicating that the solution approaches the endemic equilibrium over time. This result confirms the theoretical findings in \cite{BehRE_jmb}.

\begin{figure}[!t]
\centering
\includegraphics[scale=0.70]{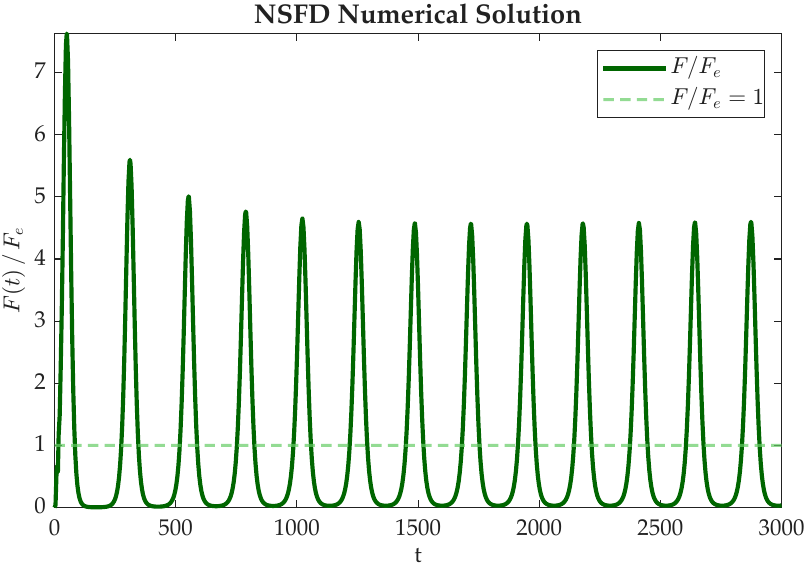}
\caption{NSFD numerical solution of problem \eqref{mod_discreto_short}--\eqref{eq:Test trap beh param} for $h=10^{-1}$. The infectivity function and the memory kernel are set as detailed in \eqref{eq:infect funct kernel} and \eqref{eq: memory kernel n2}, respectively.}
\label{fig:Test3_Paper_JMB_beh}
\end{figure}

As a second numerical experiment, we consider instead the trapezoidal infectivity function given by
\begin{equation}\label{eq:infect funct kernel}
   A(\tau)=
\begin{cases}
p_0 \dfrac{\tau-\tau_a}{(\tau_b-\tau_a)}, & \tau_a < \tau < \tau_b, \\[8pt]
p_0, & \tau_b < \tau < \tau_c, \\[8pt]
p_0 \dfrac{\tau_d-\tau}{(\tau_d-\tau_c)}, & \tau_c < \tau < \tau_d, \\[8pt]
0, & \text{otherwise},
\end{cases}
\end{equation}
with $ A_{\mu}(t) = e^{-\mu t}A(t).$
We assume $(\tau_a, \tau_b, \tau_c, \tau_d) = (4,7,11,14)$, where the parameter $p_0$ represents the weight of contacts between infected and susceptible individuals. Additional details on the infectivity function \eqref{eq:infect funct kernel} can be found in \cite{Aldis2005}. We fix $R_0 = 3.3$, and the parameter $p_0$ is determined so as to satisfy the condition defining the basic reproduction number reported in Table \ref{tab:Claudia}.
For this experiment, we adopt a different memory kernel, namely
\begin{equation}\label{eq: memory kernel n2}
    K(t) = a^2 t e^{-a t},
\end{equation}
and the parameters are chosen as
\begin{equation}\label{eq:Test trap beh param}
       T = 3000, \quad N = 5\cdot10^7, \quad \quad\mu=\frac{1}{75\cdot 365}d^{-1}, \quad S_0 = 0.7 N, \quad \alpha=5\cdot10^4.
\end{equation}
The corresponding results are displayed in Figure \ref{fig:Test3_Paper_JMB_beh}. As discussed in more detail in~\cite{BehRE_jmb}, in this case no convergence to the endemic equilibrium is observed. Instead, self-sustained oscillations arise, suggesting that the endemic equilibrium is numerically unstable. 

\section{Conclusions}\label{sec:Conclusions}
In this work, we presented EPITIME, a computational framework for the simulation of two classes of integral epidemic models, namely age-of-infection and information-dependent behavioural models. The proposed approach combines structure-preserving NSFD discretizations with modular implementations in MATLAB and Python, with the objective of providing reliable and accessible tools for the numerical study of renewal-type epidemic dynamics.

A central aspect of the framework is the use of numerical schemes specifically designed to preserve relevant qualitative properties of the continuous models, including positivity, boundedness and correct long-time behaviour. The software package includes dedicated numerical solver modules, input-validation routines, performance indicators, and a graphical user interface, which support reproducible simulation workflows and user-oriented experimentation. The numerical experiments also showed that the proposed framework can support inverse-problem settings, as illustrated by the reconstruction of an infectivity kernel from COVID-19 incidence data.

Possible directions for future work include the extension of the framework to more general epidemic kernels and further refinements of the underlying software architecture. From a modelling perspective, future developments will focus on extending the software package to  the renewal model introduced in \cite{Elefante}, which encompasses the AoI model \eqref{eq:AoI_Infectivity} as a particular case, as well as several extensions accounting for infection spread in heterogeneous populations, the distinction between symptomatic and asymptomatic individuals and viral shedding effects. A further natural extension concerns the incorporation of vaccination dynamics. In particular, vaccine uptake could be coupled with information-dependent behavioural responses and memory effects, while the renewal formulation could account for vaccination-induced changes in susceptibility and, possibly, infectivity. This would allow one to investigate, within a unified integral framework, vaccine-related behavioural feedbacks, imperfect protection, and waning immunity; see, e.g., \cite{DonofrioManfrediSalinelli2007,BuonomoDellaMarca2019,WangEtAl2016}. From a computational perspective, future versions of the GUI will include context-sensitive help menus and a compact help section or button with user instructions.

\section*{Statements}

\subsection*{Acknowledgements}
This work has been performed under the auspices of the Italian National Group for Scientific Computing (GNCS) and the Italian National Group for Mathematical Physics (GNFM) of the National Institute for Advanced Mathematics (INdAM).

\subsection*{Funding}
The authors received no specific funding for this study.

\subsection*{Author Contributions}
The authors confirm contribution to the paper as follows: Conceptualization, Bruno Buonomo and Eleonora Messina;
methodology, Eleonora Messina, Mario Pezzella and Gaetano Zanghirati; software, Claudia Panico, Mario Pezzella and Gaetano Zanghirati;
validation, Bruno Buonomo, Eleonora Messina, Claudia Panico, Mario Pezzella and Gaetano Zanghirati;
formal analysis, Claudia Panico, Mario Pezzella and Gaetano Zanghirati; investigation,  Eleonora Messina, Claudia Panico, Mario Pezzella and Gaetano Zanghirati;
data curation, Mario Pezzella; writing---original draft preparation, Claudia Panico, Mario Pezzella and Gaetano Zanghirati;
writing---review and editing, Bruno Buonomo, Eleonora Messina, Mario Pezzella and Gaetano Zanghirati; visualization, Claudia Panico, Mario Pezzella and Gaetano Zanghirati;
supervision, Bruno Buonomo, Eleonora Messina,  and Gaetano Zanghirati. All authors reviewed the results and approved the final version of the manuscript.

\subsection*{Availability of Data and Materials}
The data that support the findings of this study are openly available at \cite{IlSole24Ore2025} and \cite{EurostatPopulation2025}.
The codes of the EPITIME package, specifically developed for this work, are freely available at \href{https://github.com/ghitan/EPITIME}{github.com/ghitan/EPITIME}.

\subsection*{Ethics Approval}
Not applicable.

\subsection*{Conflicts of Interest}
The authors declare no conflicts of interest to report regarding the present study.

\subsection*{Abbreviations}
The following abbreviations are used in this manuscript: \\
\noindent 
\begin{tabular}{@{}ll}
AoI  & Age of Infection\\
CDLab & Computational Dynamics Laboratory \\
COVID-19 & COrona VIrus Disease-(20)19 \\
EPITIME & EPidemic Integral models TIMe-profile Explorer \\
FoI  & Force of Infection \\
GUI & Graphical User Interface \\
MCB & MATLAB Code Block \\
NSFD & Non-Standard Finite Difference \\
PCB & Python Code Block \\
SEIR & Susceptible, Exposed, Infected, Recovered model \\
SIR & Susceptible, Infected, Recovered model \\
\end{tabular}


\end{document}